\documentclass[journal=jacsat,manuscript=article]{achemso}
\setkeys{acs}{articletitle=true}
\setkeys{acs}{maxauthors=0}
\graphicspath{{./Figs/}}
\usepackage{amsfonts}
\usepackage{amsmath}
\usepackage{xcolor}
\usepackage{amssymb}
\usepackage[english]{babel}
\usepackage[latin1]{inputenc}
\usepackage{graphicx}
\usepackage{multirow}
\usepackage{subcaption}

\author{Sabrina M. Gericke}
\affiliation{Division of Combustion Physics, Lund University, 221 00  Lund, Sweden.}
\altaffiliation{Contributed equally to this work}
\email{sabrina_maria.gericke@forbrf.lth.se}
\author{Minttu M. Kauppinen}
\affiliation{Department of Physics and Competence Centre for Catalysis, Chalmers University of Technology,
412 96 G\"oteborg, Sweden.}
\altaffiliation{Contributed equally to this work}
\email{minttu.m.kauppinen@jyu.fi}
\author{Margareta Wagner}
\affiliation{Institute of Applied Physics, TU Wien, 1040 Vienna, Austria.}
\author{Michele Riva}
\affiliation{Institute of Applied Physics, TU Wien, 1040 Vienna, Austria.}
\author{Giada Franceschi}
\affiliation{Institute of Applied Physics, TU Wien, 1040 Vienna, Austria.}
\author{Alvaro Posada-Borb\'on}
\affiliation{Department of Physics and Competence Centre for Catalysis, Chalmers University of Technology,
412 96 G\"oteborg, Sweden.}
\author{Lisa R\"amisch}
\affiliation{Division of Combustion Physics, Lund University, 221 00  Lund, Sweden.}
\author{Sebastian Pfaff}
\affiliation{Division of Combustion Physics, Lund University, 221 00  Lund, Sweden.}
\author{Erik Rheinfrank}
\affiliation{Institute of Applied Physics, TU Wien, 1040 Vienna, Austria.}
\author{Alexander M. Imre}
\affiliation{Institute of Applied Physics, TU Wien, 1040 Vienna, Austria.}
\author{Alexei B. Preobrajenski}
\affiliation{MAX IV Laboratory, Lund University, 221 00 Lund, Sweden.}
\author{Stephan Appelfeller}
\affiliation{MAX IV Laboratory, Lund University, 221 00 Lund, Sweden.}
\author{Sara Blomberg}
\affiliation{Department of Chemical Engineering, Lund University, 221 00 Lund, Sweden.}
\author{Lindsay R. Merte}
\affiliation{Department of Materials Science and Applied Mathematics, Malm\"o University, 205 06 Malm\"o, Sweden.}
\author{Johan Zetterberg}
\affiliation{Division of Combustion Physics, Lund University, 221 00  Lund, Sweden.}
\author{Ulrike Diebold}
\affiliation{Institute of Applied Physics, TU Wien, 1040 Vienna, Austria.}
\author{Henrik Gr\"onbeck}
\affiliation{Department of Physics and Competence Centre for Catalysis, Chalmers University of Technology,
412 96 G\"oteborg, Sweden.}
\author{Edvin Lundgren}
\affiliation{Division of Synchrotron Radiation Research, Lund University, 221 00 Lund, Sweden.}

\title{\boldmath{The effect of different In$_2$O$_3$(111) surface terminations on CO$_2$ adsorption}}

\begin{document}

\section*{Keywords}
photoelectron spectroscopy, core-level shifts, heterogeneous catalysis, density functional theory, indiumoxide, CO$_2$ adsorption, methanol synthesis

\pagebreak

\section{Abstract}

In$_2$O$_3$-based catalysts have shown high activity and selectivity for CO$_2$ hydrogenation to methanol, however the origin of the high performance of In$_2$O$_3$ is still unclear. To elucidate the initial steps of CO$_2$ hydrogenation over In$_2$O$_3$, we have combined X-ray Photoelectron Spectroscopy (XPS) and Density Functional Theory (DFT) calculations to study the adsorption of CO$_2$ on the In$_2$O$_3$(111) crystalline surface with different terminations, namely the stoichiometric, the reduced, and the hydroxylated surface, respectively. The combined approach confirms that the reduction of the surface results in the formation of In ad-atoms and that water dissociates on the surface at room temperature. A comparison of the experimental spectra and the computed core-level-shifts (using methanol and formic acid as benchmark molecules) suggests that CO$_2$ adsorbs as a carbonate on all surface terminations. We find that CO$_2$ adsorption is hindered by hydroxyl groups on the hydroxylated surface.

\section{Introduction}

The emission of greenhouse gases to the atmosphere has been identified as the origin of climate change. \cite{Zimov2006} CO$_2$ has been recognized as one of the main contributors to the greenhouse effect. One suggestion to mitigate its environmental impact is to capture CO$_2$ from the atmosphere, which introduces the challenge to contain the captured CO$_2$\cite{Abello2011}. An appealing solution to this issue is the catalytical conversion of CO$_2$ into a more valuable fuel or platform chemicals. \cite{Anderson2004} The hydrogenation of CO$_2$ using renewable H$_2$ from water splitting is one promising route for the catalytical conversion of CO$_2$ to useful oxygenates such as methanol CH$_3$OH, a key building block in the chemical industry and a renewable fuel. \cite{Ye2019} Methanol can be synthesized from CO$_2$ hydrogenation by thermal catalysis \cite{Zhang2021}, by electrocatalysis \cite{Wang2022,Pan2021} by photocatalysis. \cite{Ganji2023} In thermal catalysis, Cu--ZnO--Al$_2$O$_3$ catalysts are commonly used for methanol synthesis. These Cu-ZnO-Al$_2$O$_3$ catalysts suffer, however, from deactivation, due to thermally induced sintering, \cite{Twigg2003} agglomeration of ZnO species, and oxidation of metallic Cu. \cite{Liang2019} 

In$_2$O$_3$-based catalysts have been suggested as an alternative to Cu--ZnO catalysts. Recent investigations of ZrO$_2$-supported In$_2$O$_3$ catalysts revealed high stability for CO$_2$ hydrogenation under industrial relevant conditions ($T$ = 473--573~K, $p$ = 1.0--5.0~MPa), as well as high activity and 100\% selectivity for methanol synthesis. \cite{Martin2016} The origin of the high performance of ZrO$_2$--supported In$_2$O$_3$ and the nature of the active sites of the catalysts have been the subject of intense investigations. Martin et al.\ proposed \cite{Martin2016} that the high performance originates from the high concentration of oxygen vacancies in the In$_2$O$_3$. However, these surface oxygen vacancies do not exist on stoichiometric In$_2$O$_3$(111) under ultra high vacuum (UHV) conditions and could not be created by thermal reduction.~\cite{Wagner_2014} Additionally, kinetic modeling based on density-functional theory (DFT) suggests that oxygen vacancies are not crucial for the reaction but instead that a surface structure that allows for changes in the oxidation state of the In cations is needed. \cite{Alvaro_ACS_21} Furthermore, the quantification of oxygen vacancies is based on the appearance of an additional peak at higher binding energies in the O~1s X-ray Photoelectron Spectroscopy (XPS) spectra. However, previous and current calculations on the O~1s core-level shifts of In$_2$O$_3$ surfaces show that those peaks should be assigned to hydroxyl groups rather than oxygen vacancies.\cite{Alvaro_PCCP_20, POSADABORBON2021121761}

Different atomic-scale reaction pathways have been suggested for CO$_2$--hydrogenation reactions. One is known as the reverse water--gas shift reaction (RWGS), which involves the dissociation of CO$_2$ and hydrogenation to methanol via a formyl (HCO) intermediate, whereas a more direct reaction pathway via the formation of formate (HCOO$^-$) has also been discussed in the literature. \cite{Ye2019} The high selectivity of In$_2$O$_3$-based catalysts has been attributed to the suppression of the RWGS reaction,\cite{Bielz2011} while the origin of RWGS suppression remains a subject of debate. Additionally, the effect of water on CO$_2$ hydrogenation has been discussed, and the inhibition of the CO$_2$ hydrogenation by water has been reported \cite{Martin2016} although the underlying reason behind this effect remains unknown.

Fundamental investigations of a well--defined single crystalline In$_2$O$_3$ surface in a controlled environment could advance the understanding of active sites and the effect of water on the CO$_2$ hydrogenation. The In$_2$O$_3$(111) surface is the thermodynamically most stable surface of In$_2$O$_3$ and has, moreover, been suggested to be active for CO$_2$ hydrogenation, \cite{Frei2018} making it of interest for fundamental and detailed investigations. In this paper, we apply a combined experimental synchrotron-based XPS measurements and computational DFT calculations  of well-defined In$_2$O$_3$(111) surfaces prepared under pristine (i.e., UHV) conditions. Different surface terminations of In$_2$O$_3$(111) were investigated, \cite{Franceschi2019,Wagner_2014,Wagner2017} namely the stoichiometric, the reduced, and the hydroxylated surface. 
Indium ad-atoms are identified via the In~3d core-level on the reduced surface, and OH--groups are identified via the O~1s core-level on the hydroxylated surface. The XPS fingerprints of these different surface terminations will facilitate the interpretation of future experiments performed at higher pressures. This work focuses on the adsorption of CO$_2$ on stoichiometric, reduced and hydroxylated surfaces under UHV conditions. We demonstrate that CO$_2$ forms carbonates with lattice oxygen atoms on all three surface terminations but the presence of OH groups limits CO$_2$ adsorption. Additionally, we investigated the adsorption of methanol and formic acid on the stoichiometric surface. 

\section{Experimental and Computational Methods}

The In$_2$O$_3$(111) films of 200 nm thickness were grown on yttria--stabilized zirconia by pulsed--laser deposition in Vienna, as  described in the literature. \cite{Franceschi2019} The films are single crystalline and exhibit atomically flat surfaces that can be prepared to exhibit different terminations following previously reported UHV treatments. \cite{Wagner_2014,Wagner2017} The stoichiometric surface was prepared by gentle sputtering and subsequent annealing to 800 K in 2$\times$10$^{-6}$ mbar O$_2$ for 20 min and cooling in O$_2$. The reduced In$_2$O$_3$(111) surface was obtained by annealing the stoichiometric In$_2$O$_3$(111) in UHV at 720 K for 30 minutes. The hydroxylated surface was prepared by exposing the stoichiometric In$_2$O$_3$(111) to 1 Langmuir (1\,L = 1.33$\times$10$^{-6}$\,mbar~s) of H$_2$O at room temperature. 
Prior to the adsorption experiments with methanol and formic acid, the liquids were cleaned by a freeze--pump--thaw-cycle which was repeated three times. The stoichiometric surface was flashed to 475 K to desorb any OH-groups from the surface. Subsequently, the sample was cooled down to room temperature. Once the surface had reached room temperature, 1 Langmuir of methanol or formic acid were dosed, respectively, through a leak valve with a pressure of 5$\times$10$^{-9}$ mbar.

The XPS measurements were performed at the Surface Materials Science (SMS) branch of the FlexPES beamline at the MAX IV synchrotron. \cite{Preobrajenski2023} This beamline is dedicated to high-resolution XPS and soft X-ray absorption measurements. The endstation is equipped with a Scienta DA-30 L analyzer and a preparation chamber with a low--energy electron diffraction (LEED) setup. We measured high-resolution XPS spectra of In~3d$_{5/2}$ and O~1s with an excitation energy of 600~eV and C~1s at an energy of 400~eV to ensure high surface-sensitivity. All In~3d$_{5/2}$ and the O~1s spectra were recorded with a pass energy of 20~eV and all C~1s with a pass energy of 50~eV. The spectra were recorded with an energy step size of 50~meV and the binding energy was calibrated on the valence band maximum (VBM) by setting it to 3.3~eV to compensate for band bending effects. \cite{Zhang2012}

We observed a minor potassium contamination on the sample, which accumulated on the surface when the sample was annealed. The contamination could be reduced by sputtering but not entirely removed, since it returned with annealing. The amount of potassium on the surface was estimated from the C~1s and K~2p XPS spectra. Based on the peak area of the spectra and the photoionization cross section, the potassium coverage is approximately 8\% of the saturation methanol coverage which corresponds to 3 carbon atoms per unit cell. This means approximately 0.06 potassium atoms per unit cell or $\approx0.05$~at./nm$^2$. No other contaminants could be detected within the resolution limit of XPS. The ordering of the surface was ascertained by the presence of sharp LEED spots (see Fig. S1 in the SI).

The fitting of the core--level spectra was performed using the CasaXPS software package version 2.3.24. \cite{fairley2009casaxps} A Shirley background was used for the In~3d$_{5/2}$ spectra and a linear background was used for the O~1s and the C~1s. The peak shapes that were used for the fitting are sum of a Gaussian and Lorentzian ``SGL(p)", and an asymmetric Lorentzian lineshape with tail damping ``LA($\alpha$, $\beta$, m)". Details on the fit functions can be found in the CasaXPS handbook. \cite{fairley2009casaxps}

The Vienna Ab initio Simulation Package (VASP, version 5.4.4)\cite{Kresse1993,Kresse1994a,Kresse1996a,Kresse1996} was used to perform DFT calculations with the Perdew-Buke-Ernzerhof (PBE)\cite{PBE1,PBE2} and the Heyd-Scuseria-Ernzerhof (HSE06)\cite{HSE1,HSE2,HSE3} functionals. The PBE functional was employed for all structure relaxations and O~1s and In~3d core-level shifts, whereas HSE06 was used in the CO$_2$ adsorption calculations and the C~1s core-level spectra (see SI for details). The projector-augmented wave (PAW) method was used to describe the interaction between the core and the valence-electrons\cite{Kresse1999} together with a plane-wave basis set with a $500\,\text{eV}$ cutoff energy to expand the Kohn-Sham orbitals. The valence was chosen to be 1s$^1$, 2s$^2$2p$^2$, 2s$^2$2p$^4$, and 4d$^{10}$5s$^2$5p$^1$ for H, C, O and In, respectively. The optimised bixbyite bulk structure for In$_2$O$_3$ was obtained from our earlier work.\cite{Alvaro_PCCP_20} The In$_2$O$_3$(111) surface was modelled with a 1$\times$1 surface cell of a thickness of five tri-layers (for surface termination studies and O~1s~/~In~3d core level shifts CLS) and three tri-layers (saturation coverage calculations and all hybrid calculations), with two or one bottom layers fixed at the optimised bulk positions, respectively. A $3 \times 3 \times 1$ Monkhorst--Pack-mesh was used to sample the Brillouin zone for the PBE calculations, whereas the hybrid calculations were performed using the Gamma--point approximation. The core-level shift calculations included both initial and final state effects.\cite{PhysRevLett.71.2338, FS_CLS_matensson, FS_CLS_henrik} The O~1s and In~3d shifts were computed as the difference in the energy of the system with a core hole on the atom of interest and the energy of the system with a core hole in a reference atom in the center of the slab, representing bulk In$_2$O$_3$. The C~1s shifts were computed as the difference in the energy of the system with a core hole on the atom of interest and the energy of the system with a core hole in the carbon atom of a methoxy (OCH$_3$) group placed in the same unit cell. To create the core holes, PAW potentials with one removed 3d (1s) electron were used for the In~3d (O~1s~/~C~1s) shifts. The  charge neutrality of the computational cell was maintained by employing a jellium background.\cite{ESCA,VdBossche_JCP_14} Bader charges were calculated using the code developed by the Henkelman group.\cite{bader1,bader2,bader3,bader4} Differential adsorption energies, $\Delta E_{\text{diff}}$, of molecules on the surface were calculated as
\begin{equation}
   \Delta E_{\rm{diff}} =E_{\rm{In_2O_3+nX}} - E_{\rm{In_2O_3+(n-1)X}} - E_{\rm{X}}
\end{equation}
where $E_{\rm{In_2O_3+nX}}$ and $E_{\rm{In_2O_3+(n-1)X}}$ are the total energies of the In$_2$O$_3$(111) surface slab with $n$ adsorbed molecules and a slab with $n-1$ adsorbed molecules, respectively. $E_{\rm{X}}$ is the energy of the molecule in the gas phase, which were computed at Gamma point in a simulation box of 15 \AA~side length.

\section{Results}

In the following sections we present i) the characterisation of the different surface terminations of In$_2$O$_3$(111) that will be used later to investigate CO$_2$ adsorption, ii) the adsorption of possible CO$_2$ reduction reaction intermediates (formic acid and methanol) on stoichiometric In$_2$O$_3$(111), which we also use as benchmarks for C~1s CLS, and iii) the results for CO$_2$ adsorption on the stoichiometric, reduced, and hydroxylated surface terminations of In$_2$O$_3$(111), respectively. In all sections, data from both experimental XPS and computational CLS are used to explore the structure and behavior of the In$_2$O$_3$(111) surfaces. 

\subsection{XPS fingerprints of the surface terminations of In$_2$O$_3$(111)}

\begin{figure}
    \centering
    \includegraphics[width=0.5\textwidth]{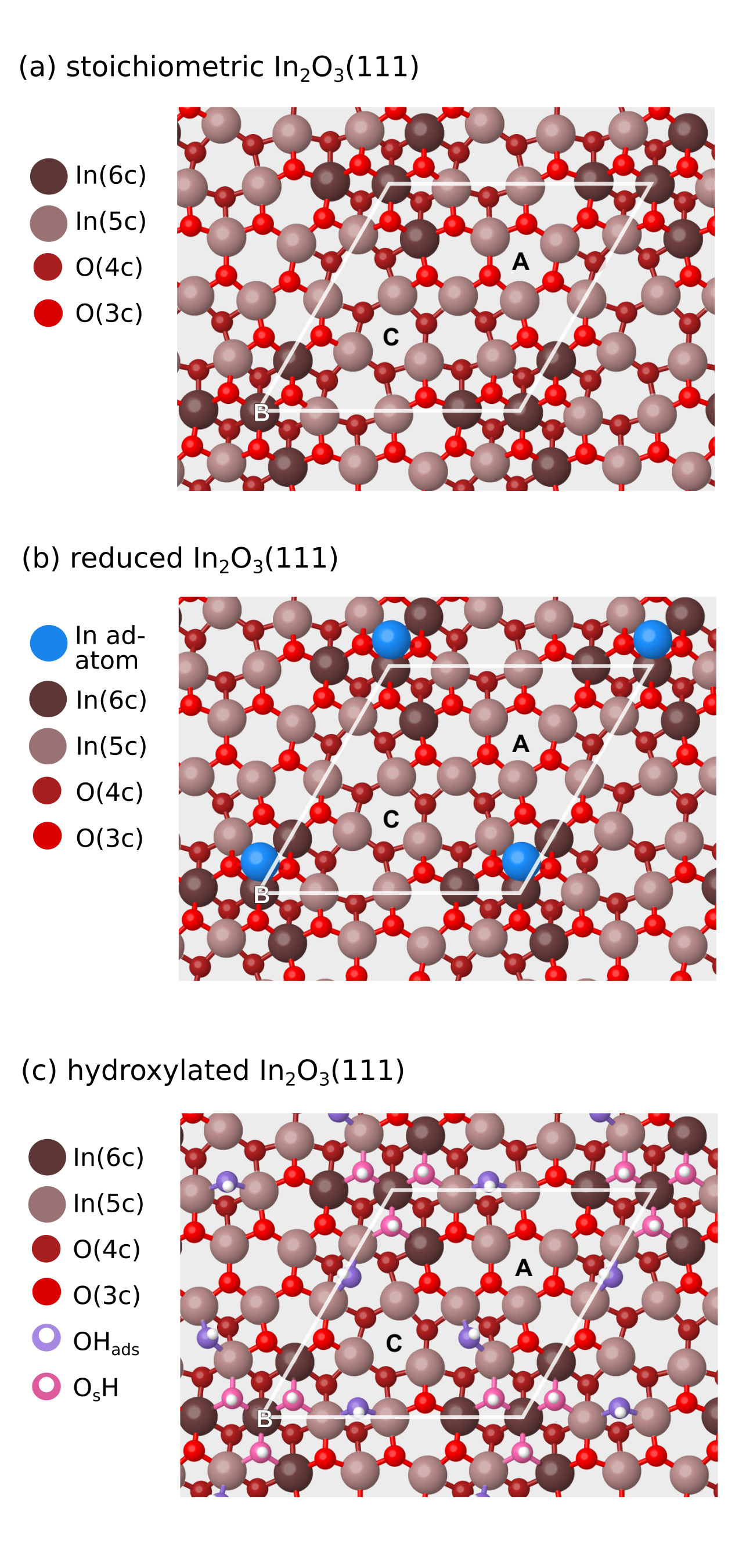}
    \caption{Top view of the first O--In--O trilayer of the surface terminations of (a) stoichiometric In$_2$O$_3$(111), (b) reduced In$_2$O$_3$(111) and (c) hydroxylated In$_2$O$_3$(111). The six- and five-fold coordinated In atoms are shown in redbrown and beige, respectively, whereas O atoms occupying sites above and below the In layer are shown in bright and dark red, respectively. On the hydroxylated surface, the O atoms of the OH groups are colored pink for O atoms belonging to the oxide lattice (O$_{\mathrm{s}}$H), and purple for the O atom originating from the dissociated water molecule (OH$_{\mathrm{ads}}$). Indium ad-atoms are colored blue. Note that the configuration depicted is the two-fold coordinated ad-atom, which is iso-energetic with the structure depicted in ref. 13. The surface cell is indicated with white lines.}
    \label{fig:terminations_dft}
\end{figure}

\begin{figure}
    \centering
        \includegraphics[width=0.6\textwidth]{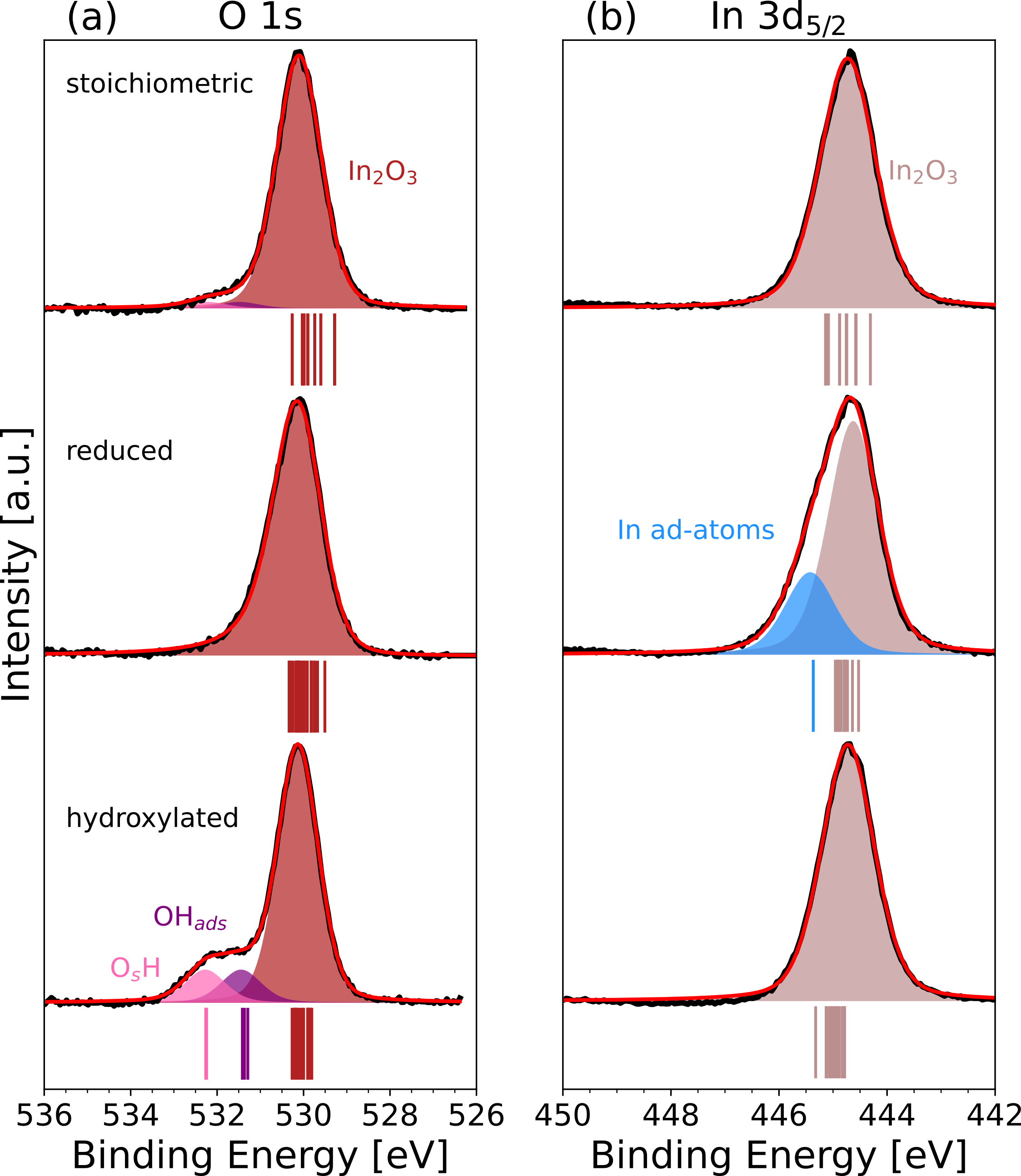}
    \caption{Panels (a) and (b) show the experimental O~1s and In~3d$_{5/2}$ core-level spectra of the three different In$_2$O$_3$ surface terminations. The spectra were background subtracted and normalized to the intensity of the main peak. Calculated CLS are reported as vertical lines below experimental spectra, and the color coding is the same as in the corresponding structures (Figure \ref{fig:terminations_dft}).}
    \label{fig:surface-Terminations} 
\end{figure}

Figures \ref{fig:terminations_dft}(a), \ref{fig:terminations_dft}(b) and \ref{fig:terminations_dft}(c) show the atomic structures of the stoichiometric, reduced and hydroxylated In$_2$O$_3$(111) surface terminations, respectively, as determined in our DFT calculations. The structures agree with previous studies of these surface terminations. \cite{Wagner_2014,Wagner2017,Franceschi2019} The experimental preparation of the surface terminations is described in the methods section. Figure \ref{fig:surface-Terminations}(a) and \ref{fig:surface-Terminations}(b) show the XPS spectra of the O~1s and In~3d$_{5/2}$ for the different surface terminations. The spectra were background subtracted and normalized to the intensity of the main peak. Details on line shapes and background functions are listed in Table S1 in the SI. 

The spectra of stoichiometric In$_2$O$_3$(111) show  a main peak at 530.1~eV in the O~1s core-level, which corresponds to the In$_2$O$_3$(111), and a small contribution of residual hydroxyl-groups at higher binding energies. In~3d$_{5/2}$ core-level shows a single symmetric peak at 444.7~eV. Contributions from differently coordinated atoms, or a surface-core-level shift could not be resolved experimentally by varying the photon energy and the electron emission angle. The DFT calculations of the core-level shifts are presented below the XPS spectra. The different lines show the results for the different indium and oxygen atoms on the surface. The difference in atomic  coordination results only in small shifts of the binding energies.

Reducing In$_2$O$_3$(111) can, in principle, lead to either the formation of oxygen vacancies or indium ad-atoms. Previous STM studies on In$_2$O$_3$(111) have reported that the thermal reduction of the In$_2$O$_3$(111) surface results in the formation of an ordered array of indium ad-atoms with one ad-atom per unit.\cite{Wagner_2014} 

The experimental O~1s core-level spectrum of thermally reduced In$_2$O$_3$(111) in Figure \ref{fig:surface-Terminations} shows a slight asymmetry towards higher binding energies. This observed asymmetry in the O 1s could result either from changes in the electronic structure of the In$_2$O$_3$ surface to a more metallic nature or from the adsorption of a small number of oxygen-containing molecules from the background gas such as small amounts of water. The In~3d$_{5/2}$ shows a strong asymmetry towards higher binding energies, which originates from the formation of a new component in the XPS spectrum. The new component has a binding energy of 445.4~eV and the bulk In$_2$O$_3$ shifts by 0.1~eV to 444.6~eV due to band bending effects (see below). It is tempting to assign the component at higher binding energy to the indium ad-atoms previously observed for the reduced surface. \cite{Wagner_2014} To validate this assignment, we calculated the relative binding energy shift of indium ad-atoms on the surface. We considered In ad-atoms placed at three different 3-fold symmetric sites on In$_2$O$_3$(111) labelled A, B and C in Figure. \ref{fig:terminations_dft} The relative stabilities, Bader charges and all In~3d CLS of the In ad-atoms on these sites calculated with the PBE exchange-correlation functional are reported in Table S8 in the SI. Our calculated ad-atom stabilities are in complete agreement with previous DFT calculations performed with another implementation of the DFT equations.\cite{Wagner_2014} 

The most stable site for In ad-atoms is the B site, where the ad-atom can coordinate to three or two oxygen atoms. The structures are found to be nearly energetically degenerate (3 meV difference), suggesting that the ad-atom can move between the sites even at low temperatures. The In~3d core-level shifts were calculated for all In atoms in the first O--In--O trilayer for stoichiometric and reduced In$_2$O$_3$(111). The In~3d shifts of the pristine surface cover a range of approximately 1~eV, with six (five)-coordinated In cations having negative (positive) shifts with respect to the bulk. The indium ad-atom on the reduced surface shows a positive shift compared to the other surface indium atoms. The experimentally observed shift of $0.8$~eV for the reduced In$_2$O$_3$(111) is very close to the 0.7~eV shift calculated for the ad-atom at the two-fold coordinated sites, indicating that In ad-atoms occupy the B site when the In$_2$O$_3$(111) is reduced, which is in agreement with the previous STM and DFT study. \cite{Wagner_2014}

We observed that the reduction of the surface causes band bending at the surface, which results in binding energy shifts of all core-levels. The effect of the band bending can be quantified from the position of the valence band maximum (VBM) by calibrating the spectra to the Fermi level of a gold foil mounted next to the sample at room temperature. The VBM is at 3.0~eV for the stoichiometric and hydroxylated surface, and at 3.2~eV for the reduced surface with In ad-atoms. A downwards band bending of 0.5~eV has previously been reported for In$_2$O$_3$(001) \cite{Hagleitner2012} between the stoichiometric and the reduced In$_2$O$_3$(001). The obtained band gap for In$_2$O$_3$(111) is close to the band gap of single crystalline In$_2$O$_3$ which has been reported to be at 2.93 $\pm$ 0.15~eV and 3.02 $\pm$ 0.15~eV for the cubic bixbyite and rhombohedral polymorphs, respectively. \cite{King2009}  

In the hydroxylation experiment with H$_2$O shown in Figure \ref{fig:surface-Terminations}, a new component appears in the O~1s spectrum at higher binding energies relative to the lattice oxygen. The shoulder can be deconvoluted into two features with binding energies of 531.5~eV and 532.3~eV, which correspond to binding energy shifts of +1.3~eV and +2.1~eV, respectively.

The DFT calculations show that hydroxylation of the In$_2$O$_3$(111) by water is energetically preferred. The adsorption energy of a single water molecule is $-$0.74~eV. Upon adsorption, the water molecule can easily dissociate at the B site with a low barrier of 0.05~eV \cite{Alvaro_ACS_21}, and an exothermic reaction energy of $-$0.57~eV. Upon dissociation, two hydroxyl (OH) groups are formed on the surface, one is the OH fragment from water, which binds to the In cations on the surface OH$_{\mathrm{ads}}$, and the other is formed as the proton from water binds to an oxygen atom on the In$_2$O$_3$(111) surface O$_{\mathrm{s}}$H. There are three equivalent sites close to the B site where water can adsorb dissociatively, and the effect of coverage on the adsorption energy is modest. Further adsorption of water takes place non-dissociatively at the C site with lower binding energies compared to the dissociative adsorption to B site. The O~1s and In~3d core-level shifts were calculated for the structure containing three dissociated water molecules (Figure \ref{fig:terminations_dft}(c)). The O~1s core-level shifts (Figure \ref{fig:surface-Terminations}) show that the OH groups give rise to characteristic peaks at higher binding energies with respect to the other surface oxygen atoms. The average computed O~1s CLS is 2.2~eV for the three O$_{\mathrm{s}}$H groups and 1.3~eV for the three OH$_{\mathrm{ads}}$ groups. The computed CLS for the two types of OH groups are in excellent agreement with the experimental XPS data (1.3~eV and 2.1~eV, respectively). The calculated O~1s CLS for molecularly adsorbed H$_2$O on the In$_2$O$_3$(111) surface is over 3~eV with respect to the bulk\cite{Alvaro_PCCP_20}. The absence of a strongly shifted peak in the O~1s XPS supports the assessment that only dissociated water is present on the surface and is in agreement with the previous STM study of water on In$_2$O$_3$(111), which showed that it is possible to achieve a coverage of three water molecules per In2O3(111) unit cell at room temperature.\cite{Wagner2017,Chen2022}

\subsection{Methanol and formic acid on stoichiometric In$_2$O$_3$(111)}

\begin{figure}
    \centering
    \includegraphics[width=0.5\textwidth]{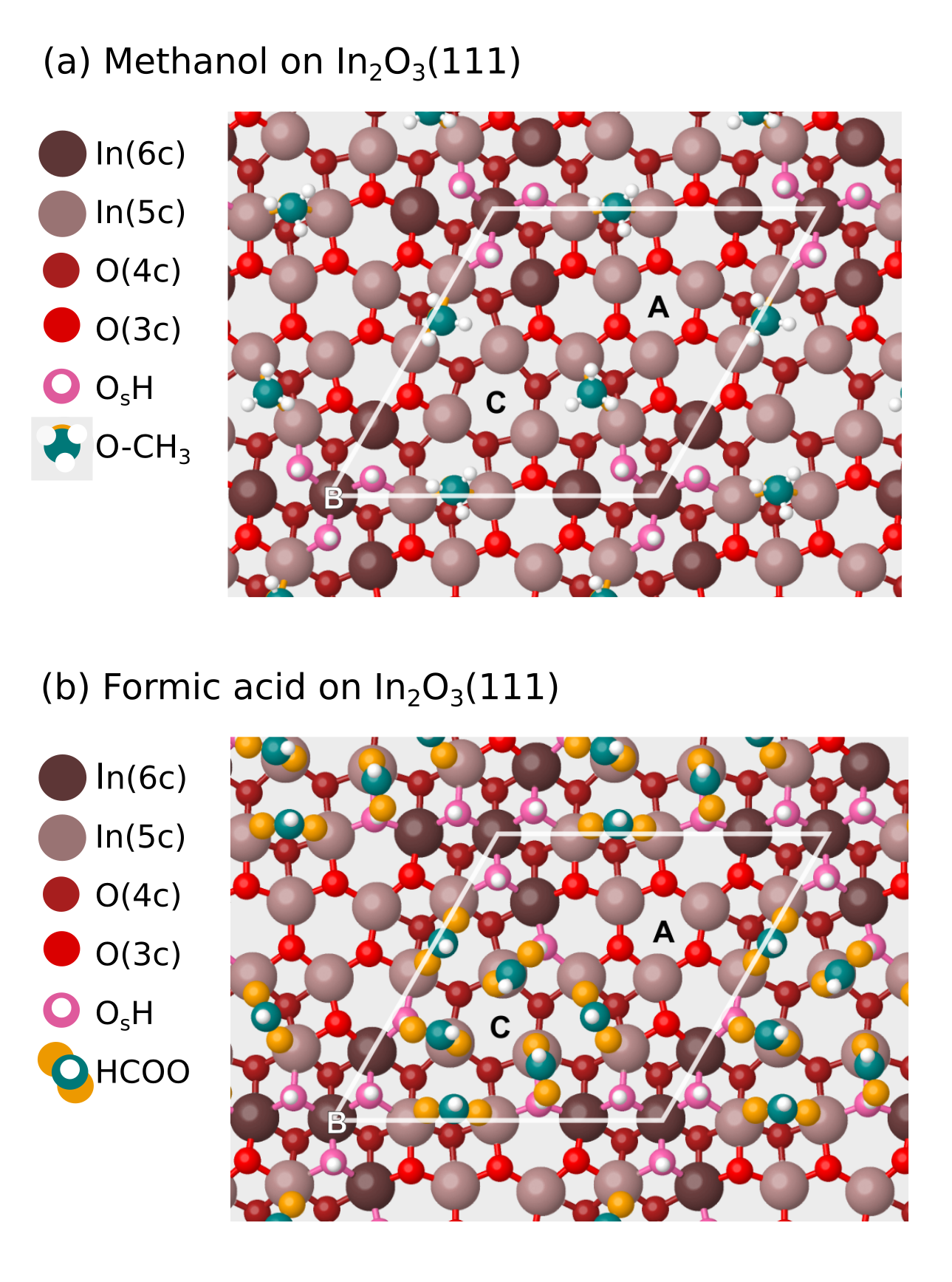}
    \caption{DFT--relaxted structures of (a) methanol and (b) formic acid on  In$_2$O$_3$(111). Carbon atoms of the HCOOH and MeOH molecules are colored teal, whereas their oxygen atoms are orange. In (a), the oxygen atom of the MeO$^-$ is hidden by the carbon atom above it.}
    \label{fig:meoh_hcooh_dft}
\end{figure}

\begin{figure}
    \centering
        \includegraphics[width=0.85\textwidth]{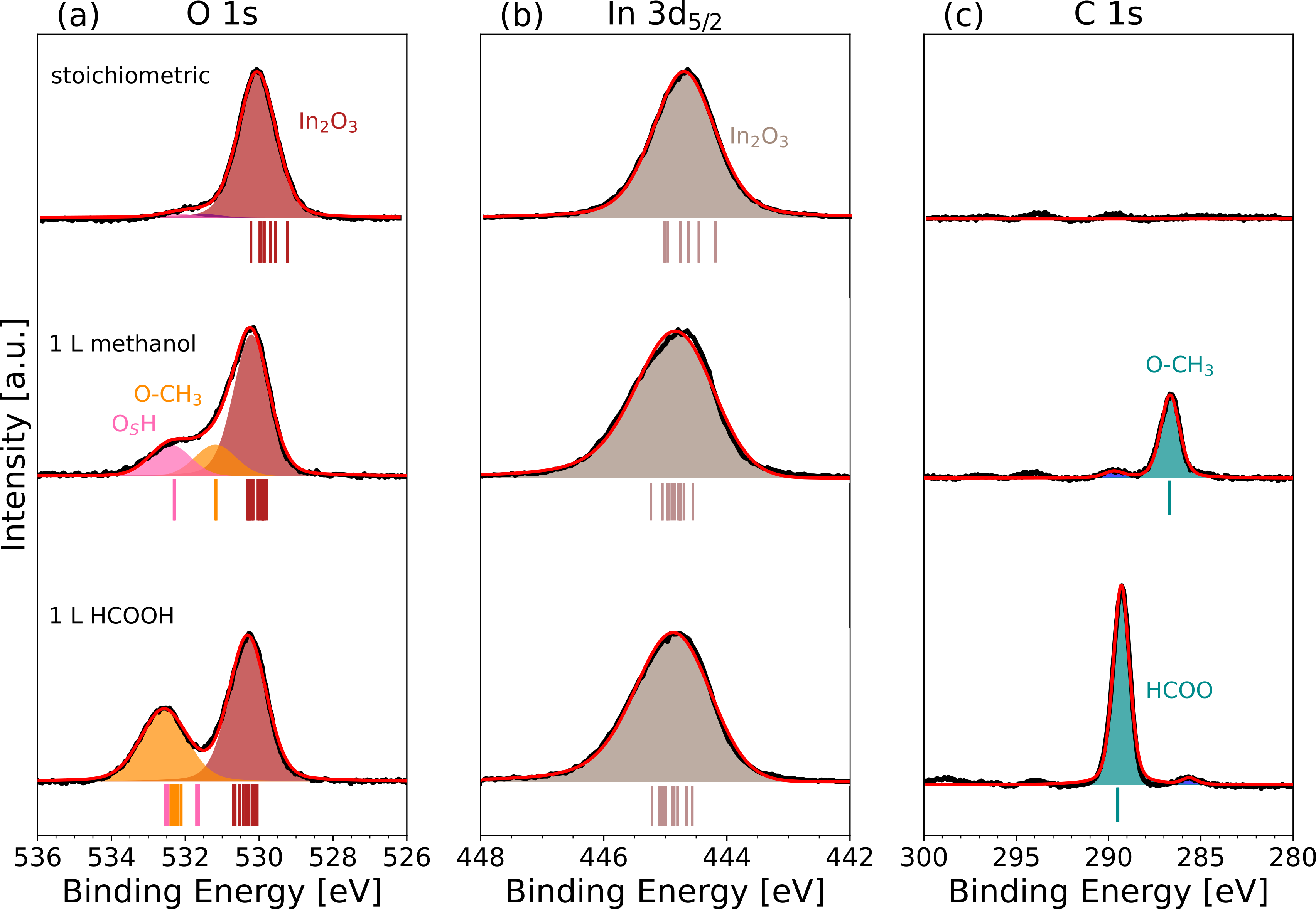}
    \caption{Panels (a), (b) and (c) show the experimental O~1s, In~3d$_{5/2}$ and C~1s core-level spectra for the stoichiometric surface and methanol and formic acid adsorbed on In$_2$O$_3$(111). Calculated CLS are indicated with vertical lines below experimental spectra. The color coding of the lines is the same as the coloring of atoms in the structural models (Figure \ref{fig:meoh_hcooh_dft}).}
    \label{fig:hydrocarbonsXPS}  
\end{figure}

To study how the methanol product and possible reaction intermediate formic acid bind to the surface, their adsorption on stoichiometric In$_2$O$_3$(111) was studied experimentally and computationally. Figure \ref{fig:meoh_hcooh_dft}(a) shows the atomic configuration of methanol on stoichiometric  In$_2$O$_3$(111). Methanol was determined to preferably adsorb dissociatively on the In$_2$O$_3$(111) forming H and O-CH$_3$ (with Bader changes of +0.63 and -0.70, respectively) pairs around the B site, preferring the same adsorption sites as dissociated water. The B site can accommodate three such pairs, which have very strong adsorption energies in the range of $-$1.2 to $-$1.0~eV. Achieving higher coverages requires the methanol to adsorb non-dissociatively around the C site, binding to an In cation through its oxygen atom. These methanol molecules have adsorption energies of only $-$0.5 to $-$0.4~eV, which suggest that only coverages of up to three methanol molecules per unit cell are achieved at room temperature. 

The O~1s, In~3d$_{5/2}$ and C~1s spectra of the methanol-covered surfaces are shown in Figures \ref{fig:hydrocarbonsXPS}(a--c) along with the pristine In$_2$O$_3$(111) for comparison. After methanol adsorption, the XPS spectra show two new peaks in the O~1s core-level at respective binding energies of 531.2~eV and 532.4~eV. These peaks can be assigned to the O--CH$_3$ fragment and the protonated oxygen atom at the B site $\text{O}_{\text{s}}\text{H}$. The shift between the peaks is calculated as 1.12~eV, which is in nice agreement with the experimental value of 1.2~eV. In the C~1s spectrum, the O--CH$_3$ groups result in a peak at 286.7~eV. The experimental In~3d$_{5/2}$ spectrum shows an asymmetry to higher binding energy after the adsorption of methanol. This is qualitatively consistent with the calculated In~3d CLS: contributions from surface In atoms are found at slightly higher binding energies than for the pristine surface. The asymmetry originates from the superposition of the bulk signal (at lower binding energies) and the signal of the surface In atoms. Interestingly, the highly coordinated indium atoms of site B give the most positive CLS on the methanol--covered surfaces, whereas on the pristine surface, they exhibit mildly negative CLS with respect to the bulk indium atoms in the middle of the slab.

Additionally, we studied the adsorption of formic acid on stoichiometric In$_2$O$_3$(111). The DFT calculations show that three formic acid molecules can adsorb dissociatively as a HCOO and H (with Bader changes of -0.75 and +0.64, respectively) pair around the B site, with the HCOO fragment in a bridging configuration between two In cations similar to the methoxy groups. The next three HCOOH molecules adsorb dissociatively around the C site: one oxygen of the HCOO fragment binds to an In cation, and the other coordinating to the H (see Figure 6 in the SI). Similarly to the case of methanol and water, adsorption around the C site is less favourable than around the B site. In contrast, the adsorption of HCOOH at site C is, however, exothermic relative to the gas phase. This indicates that In$_2$O$_3$(111) can accommodate six formic acid molecules per unit cell. 

The experimental XPS spectra are displayed in Figure \ref{fig:hydrocarbonsXPS}. The O~1s spectrum shows an additional peak at 532.6~eV with a large FWHM of 1.5~eV indicating that multiple components are contributing to this peak. After the adsorption of formic acid, the In~3d$_{5/2}$ shows an increased asymmetry, similar to the methanol case. The C~1s spectrum shows a peak at 289.3~eV. 

The computed O~1s CLS for HCOOH on the surface shows three groups of peaks. The surface oxygens that do not take part in the HCOOH adsorption show the lowest relative shifts. The $\text{O}_{\text{s}}\text{H}$ groups around site C have an average shift of 1.4~eV, whereas HCOO and $\text{O}_{\text{s}}\text{H}$ groups around site B have shifts ranging from 1.8 to 2.2~eV. The computed O~1s shifts are in fair agreement with the O~1s XPS spectrum. Similarly to methanol adsorption, the In~3d CLS show that the surface In cations are shifted to slightly higher binding energies than on the pristine surface. 

The C~1s CLS of the HCOOH were computed relative to the C~1s CLS of the reference O--CH$_3$ group (see SI for a detailed discussion). The calculations were performed by placing a dissociated methanol molecule in the same unit cell with formic acid, and calculating the total energy with a core hole on each carbon atom. The CLS were calculated for two separate cases, where the HCOO and  H pair is bound to either the B site or the C site. The HCOO-fragment at the B site has a shift of 2.78~eV, while the HCOO bound to the C site has a shift of 2.82~eV, relative to O--CH$_3$. The average relative CLS of the two HCOO-groups and the O--CH$_3$ group is close (2.8~eV) to the experimentally observed difference in the binding energy (2.6~eV) of the HCOOH and O--CH$_3$ C~1s peaks. Further analysis on the effect of the binding configuration, exchange-correlation functional, initial and final state effects, and surface coverage on calculated C~1s core-level shifts is presented in the SI.

The number of formic acid molecules per unit cell can be estimated from the area of the C~1s peak using the peak intensity of the O--CH$_3$-peak as a reference, under the assumption that this saturated surface is covered by three dissociated methoxy-molecules. Based on this assumption and the support from the DFT calculated adsorption energy trends, we can conclude that six formic acid molecules can adsorb per unit cell, as illustrated in Figure \ref{fig:meoh_hcooh_dft}(b). 

\subsection{CO$_2$ adsorption on different surface terminations of In$_2$O$_3$(111)}

\begin{figure}
    \centering
    \includegraphics[width=\textwidth]{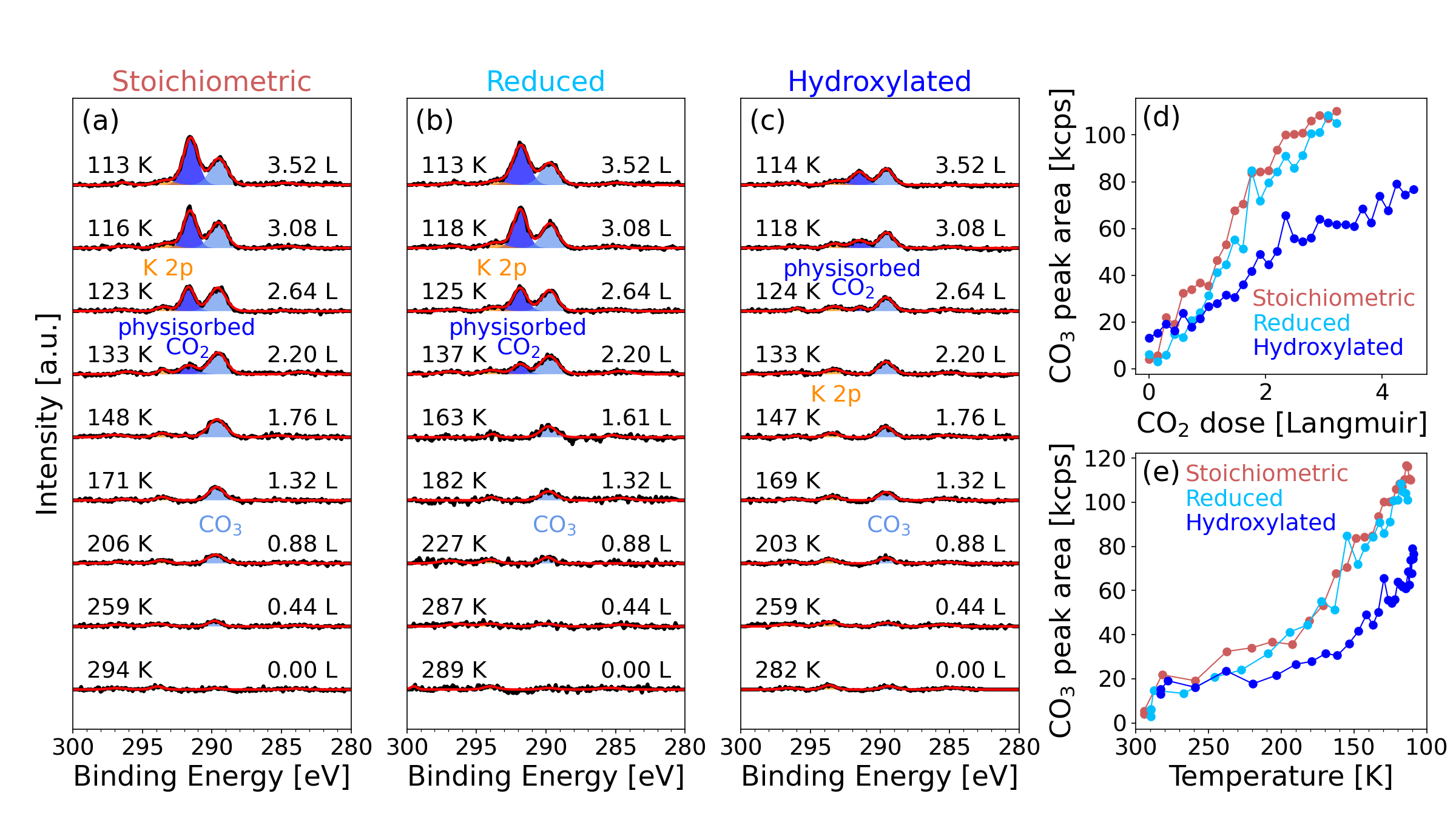}
    \caption{Measurements of CO$_2$ adsorption during sample cooling. C~1s core-level of (a) the stoichiometric surface, (b) the reduced surface and (c) the hydroxylated surface during CO$_2$ adsorption at 
    mbar CO$_2$ while the samples are cooled down from room temperature to 100 K. The numbers on the left hand side of every panel give the temperature in Kelvin as the samples are cooled down and the numbers on the right hand side of the panel gives the CO$_2$ dose in Langmuir. Panel (d) shows the area of the CO$_3$ peak as a function of CO$_2$ dose in Langmuir and panel (e) shows the CO$_3$ peak area as a function of the sample temperature.}
    \label{fig:CO2_adsorption} 
\end{figure}

The adsorption of CO$_2$  was monitored experimentally on different terminations of the In$_2$O$_3$(111) surface. The surfaces were exposed to 5$\times$10$^{-9}$ mbar CO$_2$ while being cooled from room temperature to 100 K. We chose this experimental route to minimize the adsorption of H$_2$O from the background that would otherwise occur when first cooling the sample and later dosing CO$_2$. Figure \ref{fig:CO2_adsorption}(a), \ref{fig:CO2_adsorption}(b) and \ref{fig:CO2_adsorption}(c) show the C~1s spectra of the stoichiometric, the reduced and the hydroxylated In$_2$O$_3$(111) during CO$_2$ adsorption, respectively. On all three surfaces, CO$_2$ adsorption results in the development of an XPS feature at 289.6 $\pm$ 0.1~eV. Its position is consistent with the formation of carbonate (CO$_3$). \cite{Yang2017} The peak becomes visible on all three surfaces at temperatures around 200 K and grows as the sample is cooled further. In comparison, the peak growth is slower on the hydroxylated surface than on the other two surface terminations, indicating that the hydroxyl groups on this surface hinder the adsorption of CO$_2$. Conversely, adsorption on the reduced surface proceeds in a similar way as on the stochiometric surface, showing that In ad-atoms do not affect the CO$_2$ adsorption. Figures \ref{fig:CO2_adsorption}(d) and \ref{fig:CO2_adsorption}(e) show the peak area of the carbonate peak as a function of the amount of dosed CO$_2$ and temperature, respectively. A new peak appears as the surfaces are cooled down to 140 K. This peak has a binding energy of 291.6 $\pm$ 0.1~eV, which originates from physisorbed CO$_2$. \cite{Yang2017} Again, the growth of this peak is considerably slower on the hydroxylated surface. 

\begin{figure}
    \centering
    \includegraphics[width=0.5\textwidth]{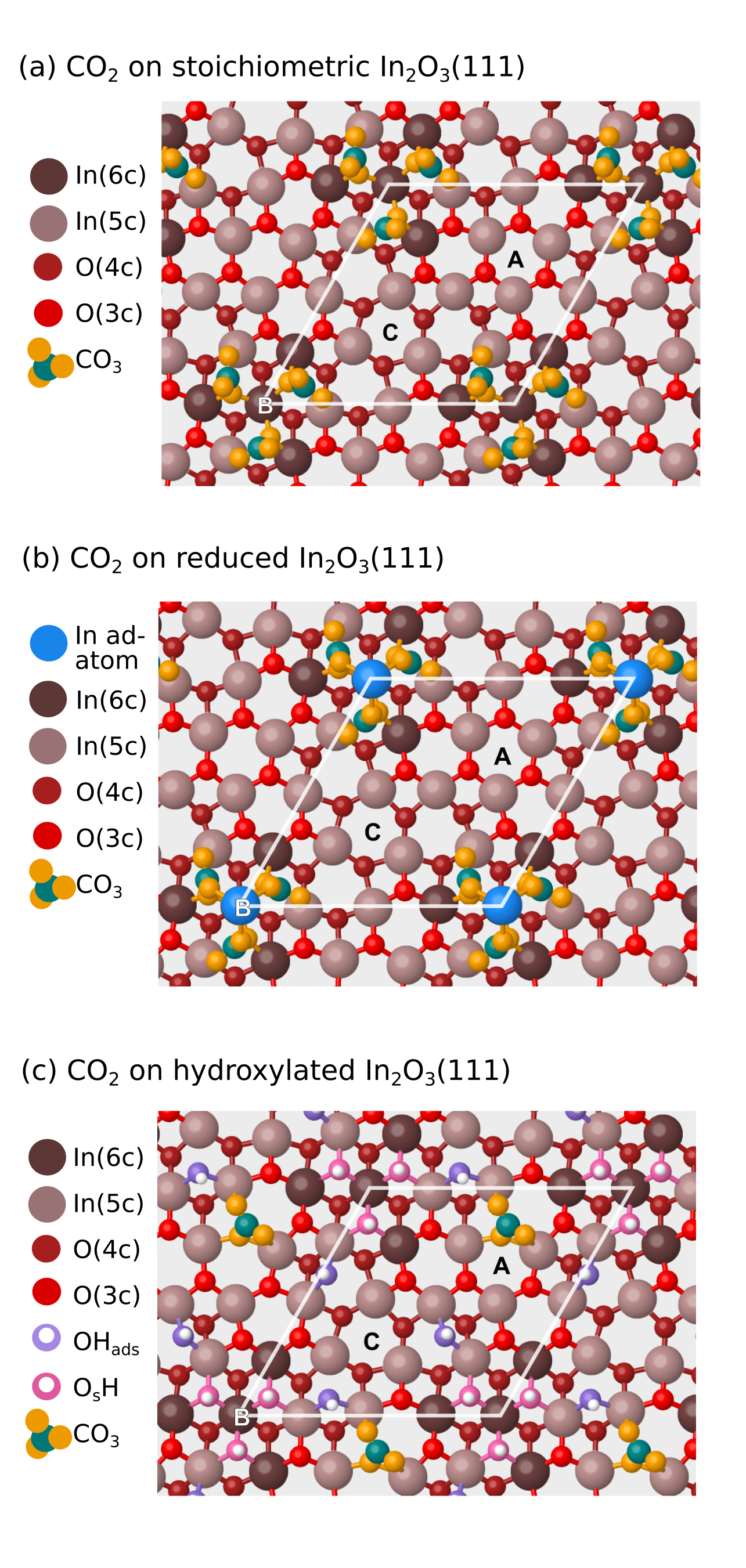}
    \caption{DFT--relaxted structures of CO$_2$ adsorption on (a) stoichiometric In$_2$O$_3$(111), (b) reduced In$_2$O$_3$(111) and (c) hydroxylated In$_2$O$_3$(111). Carbon atoms are colored teal, while the oxygen atoms forming the CO$_3$ carbonate species are colored orange.}
    \label{fig:co2_dft}
\end{figure}

We performed DFT calculations to identify the adsorption configuration of CO$_2$ on the differently terminated In$_2$O$_3$(111) surfaces. The resulting structures are illustrated in Figure \ref{fig:co2_dft}. On the pristine surface, the CO$_2$ molecule can only adsorb by binding to a surface oxygen as a bent carbonate (CO$_3$) species with a Bader charge of $-$1.46 e. The adsorption is preferred on the undercoordinated oxygen atoms around the B site (Figure \ref{fig:co2_dft}a). Here, the adsorption energy is $-$0.9~eV. The adsorption energy is lowered when more than one CO$_2$ molecule is adsorbed at the B site, and once all three oxygen atoms are occupied, additional CO$_2$ adsorbs close to site A instead. The adsorption energy of additional CO$_2$ at site A is weak ($-$0.3 to $-$0.1~eV), which indicates that at most three CO$_2$ molecules can adsorb as carbonate on the pristine In$_2$O$_3$(111) surface. The C~1s CLS of the carbonate is 2.9~eV with respect to the methoxy group peak. The relative shift is in good agreement with experiments as the difference in the binding energy of the methoxy and carbonate C~1s XPS peaks is 3.0~eV. 

On the hydroxylated surface, the oxygen atoms close to site B have all been converted to hydroxyl groups. The OH$_{\mathrm{ads}}$ and O$_{\mathrm{s}}$H pair bind stronger to the site than CO$_2$, and preferably occupy adjacent sites.\cite{Wagner2017} It is energetically unfavorable to displace a hydroxyl group and to form a carbonate, thus CO$_2$ adsorption is blocked by water at the B site. The only sites left available for CO$_2$ adsorption are the oxygen atoms around site A, where the adsorption energy of a single CO$_2$ is $-$0.7~eV. Adsorption of additional CO$_2$ is less favorable: $-$0.5 and $-$0.4~eV for the second and third molecule, respectively. Interestingly, adsorption of carbonate at site A is more favorable when site B hosts water than when it hosts other carbonate species. Unlike dissociated water, the formation of a carbonate requires a charge transfer from the surface to adsorbate, thus the In$_2$O$_3$(111) is not able to accommodate as many carbonates. 

As an alternative to adsorbing as carbonate, CO$_2$ could in principle react with one of the O$_{\mathrm{s}}$H or OH$_{\mathrm{ads}}$ groups on the hydroxylated surface to form formate (HCOO$^-$) or bicarbonate (HCO$_3^-$). The carbon atom of the formate species has a C~1s CLS of 3.0~eV with respect to the methoxy group peak (similar to the value of 2.9~eV of C~1s of carbonate species, see above), which would also be in very good agreement with the XPS data. However, adsorption is highly endothermic (approximately 3~eV) and thus unfavorable. Bicabonate species are more stable than formate, however, the calculated C~1s CLS of all considered HCO$_3$-configurations are strongly shifted, approximately 4~eV higher binding energy, with respect to the methoxy peak. Therefore, we propose that CO$_2$ adsorbs as a carbonate also on the hydroxylated surface.

On reduced In$_2$O$_3$(111), we find that CO$_2$ can adsorb as a carbonate in essentially the same geometry as on the pristine surface, additionally coordinating to the In ad-atom. A single carbonate pushes the ad-atom off-center of the B site so that it preferably occupies the two-fold coordinate site, however, addition of more CO$_2$ pushes the ad-atom to the central position. The carbonate adsorption energy is more exothermic than on the pristine surface, and it is possible to populate all three B--site oxygen atoms simultaneously.

It is possible to estimate the number of molecules per unit cell by comparing the peak area of the adsorbed CO$_3$ in the experimental C~1s spectrum to the peak area of the adsorbed methanol. With a methanol coverage of three molecules per unit cell, the CO$_2$ coverage on the stoichiometric and reduced surface corresponds to approximately 1.7 molecules per unit cell, and to one molecule per unit cell on the hydroxylated surface. While the experimentally observed coverage on the hydroxylated surface is in agreement with the DFT calculations, the experimental coverage on the stoichiometric and on the reduced surface of 1.7 molecules per unit cell is lower than the three molecules per unit cell predicted by DFT calculations. We speculate that the discrepancy is due to the adsorption of water from residual gas in the vacuum chamber as the samples are cooled down. The adsorbed water molecules effectively block adsorption sites for the CO$_2$ as they do on the hydroxylated surface and thus lower the CO$_2$ coverage observed in the experiments.

\begin{figure}
    \centering
    \includegraphics[width=0.85\textwidth]{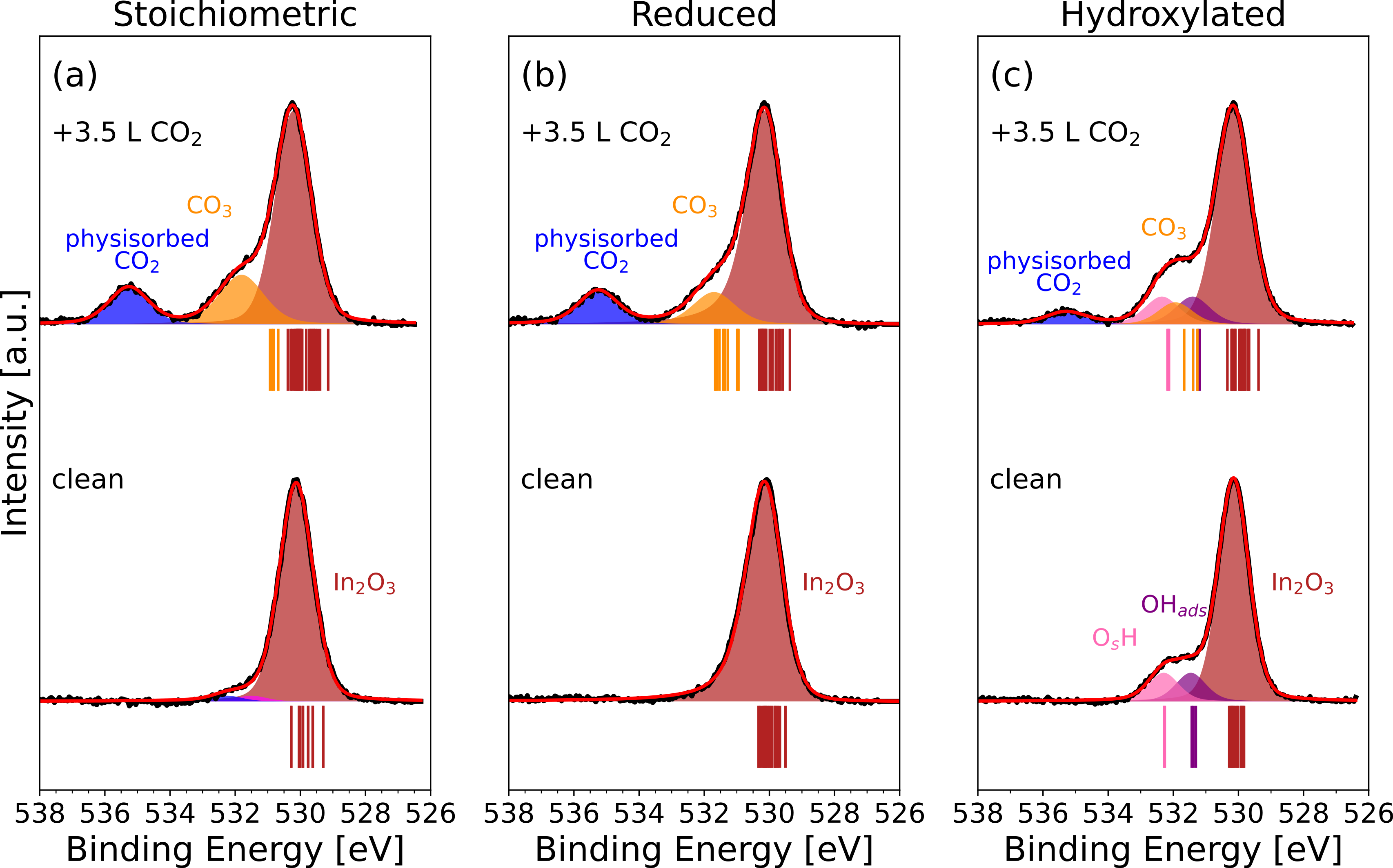}
    \caption{O~1s core-level spectra of (a) the stoichiometric, (b) the reduced and (c) the hydroxylated In$_2$O$_3$(111) surface at room temperature prior to the adsorption of CO$_2$ (bottom), and after the adsorption of 3.5 L CO$_2$ at 5$\cdot$10$^{-9}$ mbar (top) at the final temperature of approximately 100 K. The corresponding calculated CLS are indicated with vertical lines under each spectrum, their colors match the coloring of the atoms in the atomic structure figures.}
    \label{fig:O1s_CO2} 
\end{figure}

\begin{figure}
    \centering
    \includegraphics[width=0.85\textwidth]{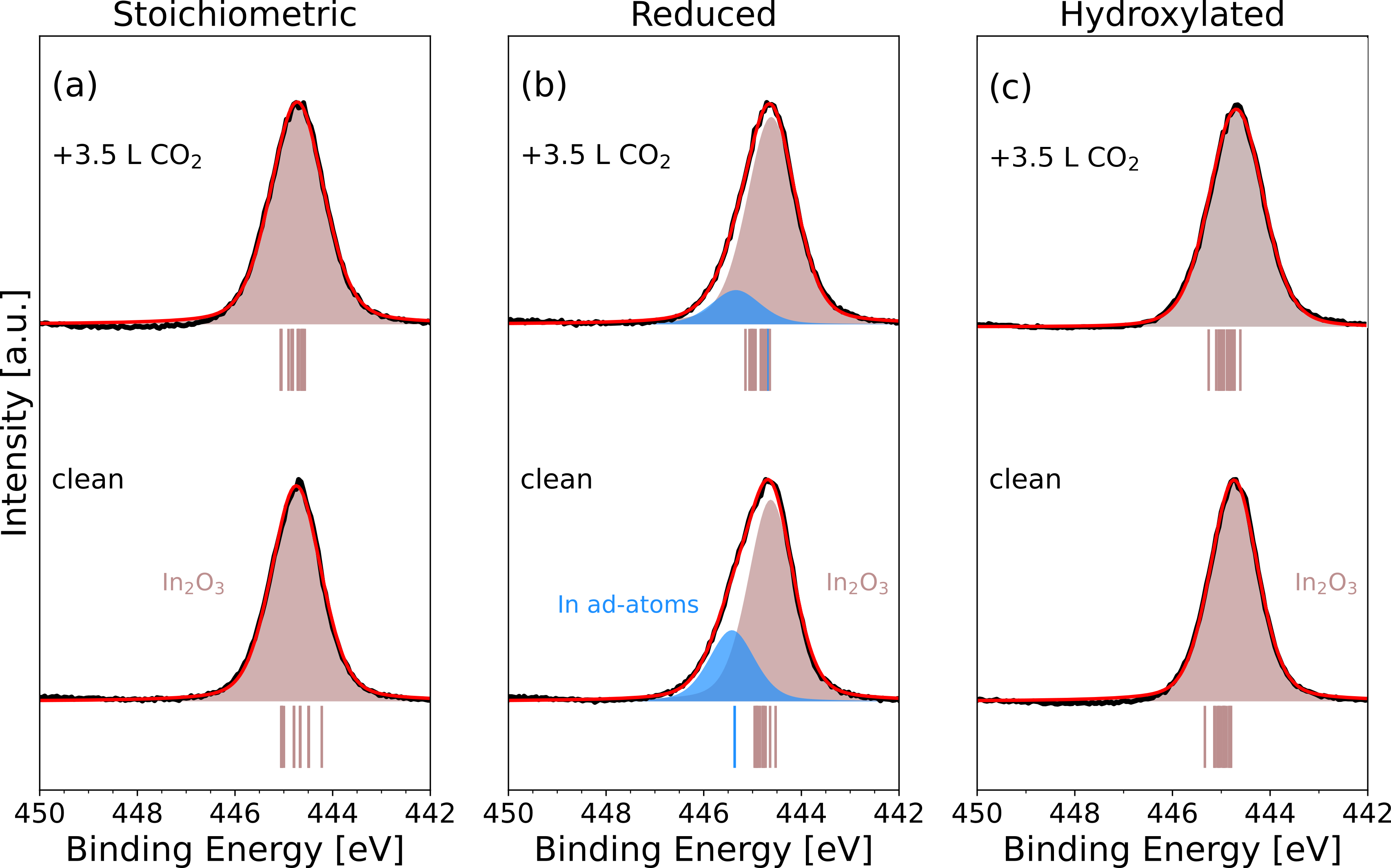}
    \caption{In~3d$_{5/2}$ core-level spectra of (a) the stoichiometric, (b) the reduced and (c) the hydroxylated In$_2$O$_3$(111) surface at room temperature prior to the adsorption of CO$_2$ (bottom), and after the adsorption of 3.5 L CO$_2$ at 5$\times$10$^{-9}$ mbar (top) at the final temperature of approximately 100 K. The corresponding calculated CLS are indicated with vertical lines under each spectrum.}
    \label{fig:In3d_CO2} 
\end{figure}

Figures \ref{fig:O1s_CO2} and \ref{fig:In3d_CO2} show the experimental XPS spectra of the O~1s and the In~3d$_{5/2}$ core-level of the three surface terminations after CO$_2$ adsorption, respectively. To support the experimental data, we also performed DFT calculations for CO$_2$ adsorbed in a carbonate configuration on the three In$_2$O$_3$(111) surface terminations. The CLS in Figures \ref{fig:O1s_CO2}  and \ref{fig:In3d_CO2} were calculated at a coverage of three carbonates per unit cell for the pristine and reduced surface, and one carbonate per unit cell for the hydroxylated surface. For the stoichiometric In$_2$O$_3$(111), a minor broadening of the  In$_2$O$_3$ peak in the O~1s and In~3d core-level is observed after the CO$_2$ adsorption. The adsorption of CO$_2$ appears to diminish the peak from the In ad-atoms by 55 \% in the In~3d core-level whereas no change is observed in the O~1s spectrum. For the hydroxylated surface, the O~1s peak and the In~3d peak broaden after CO$_2$ adsorption. The DFT calculations show that the oxygen atoms of the carbonate (CO$_3$) are positively shifted with respect to a bulk oxygen on the pristine surface, however the shifts are not as clearly distinguishable from the main surface oxygen peak as determined experimentally. On the reduced surface, the O~1s CLS of the CO$_3$ species and the surface oxygens are more separated and agree well with the experimental spectrum. Finally, on the hydroxylated surface, the CO$_3$ CLS are partially overlapping with the OH$_{\mathrm{ads}}$ peak, however the computed CLS are still in good qualitative agreement with the observed XPS. Calculated In~3d CLS show very little change upon CO$_2$ adsorption for the pristine and hydroxylated surface. On the reduced surface the CO$_2$ shifts the peak of the indium ad-atoms towards the bulk In$_2$O$_3$, agreeing very well with the experimentally observed suppression of the ad-atom peak upon CO$_2$ adsorption. 

\section{Discussion}

We have presented XPS spectra of stoichiometric, reduced and hydroxylated In$_2$O$_3$(111), as well as changes upon adsorption of methanol and formic acid adsorbed on stoichiometric In$_2$O$_3$(111). The experimentally observed XPS spectra are overall in good agreement with the presented DFT core-level shift calculations. The experimental spectra and calculated CLS are important references for experiments performed at elevated pressures and temperatures. 

Additionally, we studied the adsorption of CO$_2$ on different surface terminations of In$_2$O$_3$(111). The experiments showed that hydroxyl groups on In$_2$O$_3$(111) partially block the adsorption of CO$_2$. Water has been reported to decrease the activity of CO$_2$ hydrogenation for In$_2$O$_3$ \cite{Martin2016} and other CO$_2$ hydrogenation catalysts.\cite{Saito1996} Our previous DFT-based microkinetic models have also shown that OH can block surface sites of In$_2$O$_3$(110), which leads to a negative reaction order with respect to the partial pressure of water.\cite{Alvaro_ACS_21} Our present DFT calculations indicate that water and CO$_2$ preferably adsorb at the same surface site. The dissociative water adsorption to the site is stronger than the physisorption of CO$_2$, therefore the resulting hydroxyl groups can block CO$_2$ from adsorbing as carbonate. Our computed C~1s core-level shift for carbonate closely matches the experimental shift relative to the methanol C~1s peak. The agreement between experiment and computations was achieved by using the well defined methanol and formic acid C~1s peaks as reference. For the C~1s CLS we find that including exact exchange by employing a hybrid functional is vital to achieve the correct relative shifts for carbon-containing adsorbates on In$_2$O$_3$(111).

The adsorption of CO$_2$ on different catalysts for CO$_2$ hydrogenation such as CeO$_2$, Rh and CuZn has been studied previously. CO$_2$ has been reported to adsorb as a carbonate on CeO$_2$ as well as on Zn deposited on copper surfaces, \cite{Yang2017,Koitaya2019} while it was reported to dissociate on Rh \cite{Kim2020} and stepped Cu surfaces. \cite{Hagman2018} The present measurements show that CO$_2$ does not dissociate on In$_2$O$_3$(111) when adsorbed at a pressure of  5$\times$10$^{-9}$ mbar and temperatures below room temperature. We could not observe the formation of formate on the surface when the CO$_2$ was adsorbed on the hydroxylated surface. This suggests that higher pressures, temperatures or additional gases are required to activate the CO$_2$ for the hydrogenation reaction. Our previous DFT studies suggest that the In$_2$O$_3$ surface is partially hydrogenated under typical reaction conditions,\cite{Alvaro_PCCP_20} with kinetic studies on hydrogenated In$_2$O$_3$(110)\cite{Alvaro_ACS_21} supporting the notion that a hydrogenated surface forms the active site for the methanol synthesis from CO$_2$. In contrast to the hydrogenated surface, hydroxylation by water does not result in a change in oxidation state \cite{Alvaro_PCCP_20,C9CP04097H} for the surface In atoms, and does not facilitate the activation of CO$_2$.

A detailed understanding of the adsorption of CO$_2$ on In$_2$O$_3$ is an important  step towards understanding the catalytic process of thermal CO$_2$ hydrogenation over In$_2$O$_3$-catalysts on the atomic scale. In a previous study, the reaction mechanism of CO$_2$ hydrogenation has been attributed to the interaction of reactants with O-vacancies. \cite{Frei2018} However, we have no evidence of the existence of these defects in the processes that we have studied so far.

\begin{figure}
    \centering
    \includegraphics[width=0.5\textwidth]{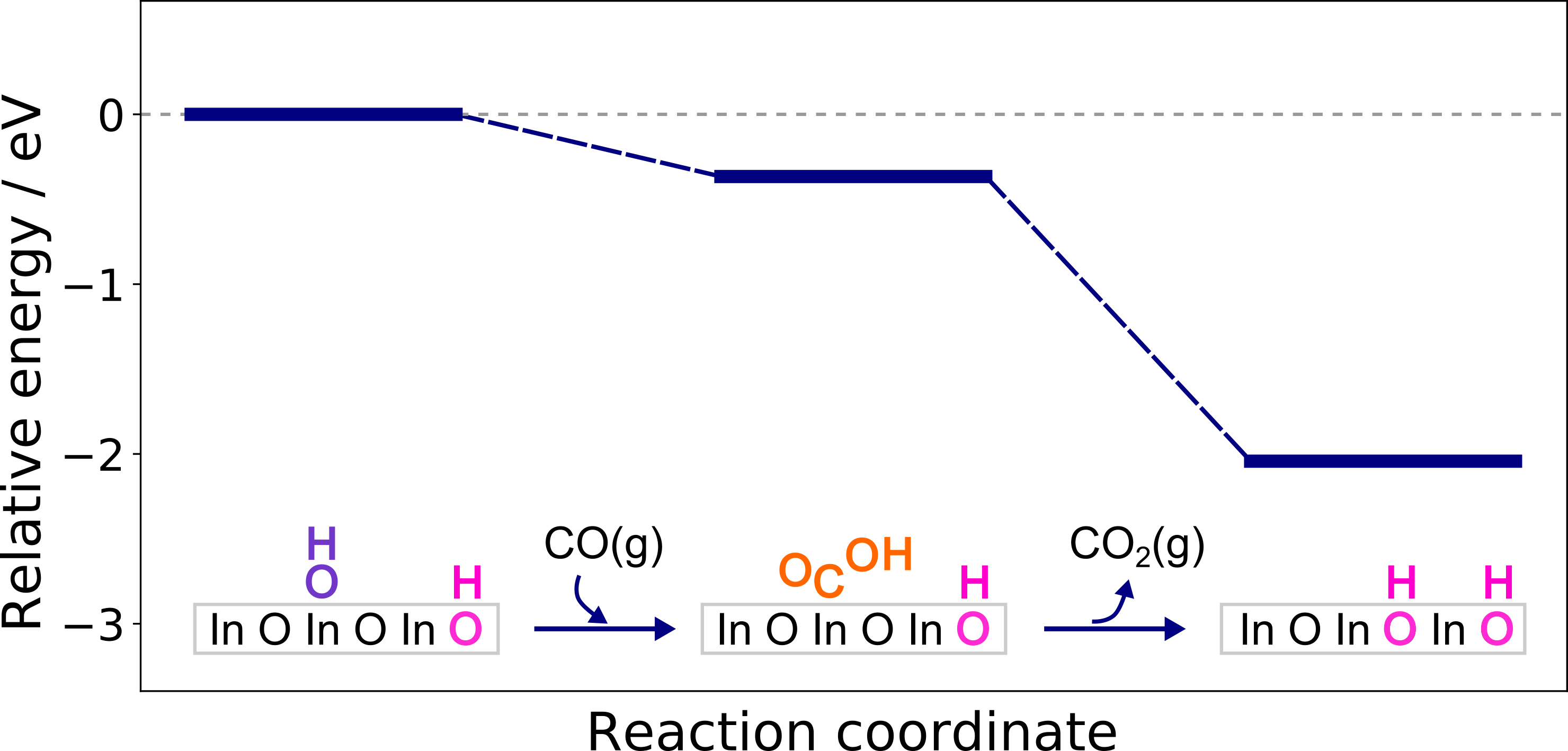}
    \caption{DFT calculated energy profile for the water-gas shift reaction between a surface O$_{\mathrm{ads}}$H group and a gas-phase CO.}
    \label{fig:PES} 
\end{figure}

The catalytic activity of In$_2$O$_3$ and CuZn increases when CO is added to the CO$_2$ and H$_2$ gas mixture.\cite{Martin2016} In this context, it is interesting to note that for CuZn catalysts it has been shown that the increased activity results from the removal of hydroxyl groups from the surface by the CO via the WGS reaction. \cite{Ruland2020} We speculate that a similar mechanism may occur for the In$_2$O$_3$ surfaces, explaining the promotional effect of adding CO to the CO$_2$ and H$_2$ gas feed. Without calculating barriers, we can evaluate the thermodynamic feasibility of the WGS reaction on In$_2$O$_3$(111) by considering the reaction between CO and a surface hydroxyl (see Figure \ref{fig:PES}). It has been shown previously in the case of \textit{m}-ZrO$_2(\overline{1}11)$ that CO and OH cannot directly form formate in a single elementary step, and instead react to form its structural isomer, carboxyl.\cite{Minttu} Thus, we also consider the WGS to proceed through carboxyl, which consequently dissociates into CO$_2$ and a proton on the surface. Our thermodynamic analysis shows that the reaction is feasible on the In$_2$O$_3$(111), therefore it may be possible that CO can remove OH groups through WGS. 

\section*{Conclusions}

We have studied different surface terminations of In$_2$O$_3$(111), stoichiometric, reduced (with In ad-atoms), and hydroxilated, and the adsorption of CO$_2$ on these different surface terminations using XPS and DFT. Our results confirm the structure of the In ad-atom overlayer and the adsorption site of the hydroxyl groups reported in the literature. The experiments on the CO$_2$ adsorption showed that the In ad-atoms do not hinder the adsorption of CO$_2$ while hydroxyl groups on the surface partially block the adsorption of CO$_2$. The DFT calculations showed that the CO$_2$ does not dissociate and adsorbs as a carbonate on all studied surface terminations of In$_2$O$_3$(111).

\section*{Conflicts of interest}
There are no conflicts to declare.

\section*{Acknowledgements}
This project is financially supported by the Knut and Alice Wallenberg  (KAW)  project ``Atomistic design of new catalysts" (project no. KAW2015.0058), the Swedish Research Council (project no. 2018-03434 and 2020-05191), the Swedish Foundation for Strategic Research (project no. ITM17-0045). The calculations were performed at PDC via a SNIC grant. This work was supported by the Austrian Science Fund (FWF), Project V 773-N (Elise-Richter-Stelle, M.W.).
We acknowledge MAX IV Laboratory for time on Beamline FlexPES under Proposal 20200070 and 20210938. Research conducted at MAX IV, a Swedish national user facility, is supported by the Swedish Research council under contract 2018-07152, the Swedish Governmental Agency for Innovation Systems under contract 2018-04969, and Formas under contract 2019-02496.
G.F. and U.D. acknowledge support from by the European Research Council (ERC) under the European Union's Horizon 2020 research and innovation program (grant agreement No. 883395, Advanced Research Grant 'WatFun'). M.R., A.I. and E.R. acknowledge support from the Austrian science fund (FWF) via the SFB project F81 ``Taming complexity in materials modeling" (TACO).

\section*{Supporting Information}

\subsubsection{XPS Fits}
Tables showing the line shapes, background shapes and FWHM of the XPS spectra shown in the main text.

\subsubsection{DFT calculations}

Comparison of stabilities, Bader charges, and In~3d core-level shifts of different ad-atom placements. Comparison of stabilities and PBE/HSE06-calculated C~1s core-level shifts of different carbon-containing species on the hydroxylated In$_2$O$_3$(111) surface. Plots of adsorption energy as a function of coverage for water, methanol, formic acid, and CO$_2$ on the pristine In$_2$O$_3$(111) surface. Plot of adsorption energies of CO$_2$ on the pristine, reduced and hydroxylated In$_2$O$_3$(111) surface as a function of CO$_2$ (carbonate) coverage.

\bibliography{refs} 

\pagebreak

\begin{figure}
    \centering
        \includegraphics[width=0.85\textwidth]{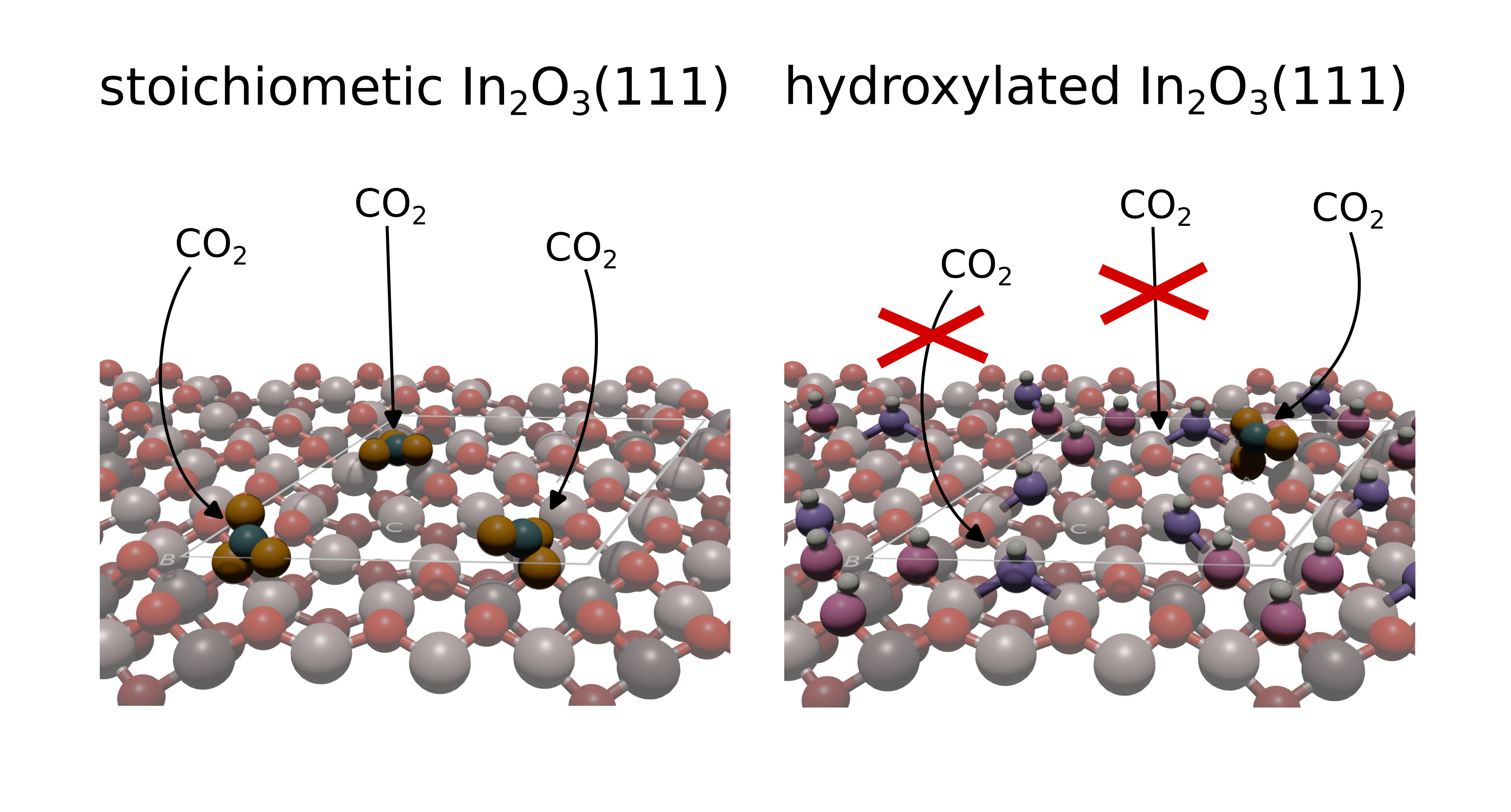}
    \caption{Table Of Contents (TOC) graphic.} 
\end{figure}

\end{document}


\section{XPS fitting parameters}

The following section contains the fitting parameters used for the XPS spectra in the main text.

\begin{table}[H]
\centering
\begin{tabular}{c cccccc}
Surface  & Core&  Peak& Binding energy & FWHM & Line & Background \\
 termination&  level & label  &  [eV] & [eV] & shape& shape\\ 
 \specialrule{2.5pt}{1pt}{1pt}
 & In 3d$_\mathrm{5/2}$  & In$_2$O$_3$ & 444.7 & 1.2 &SGL(30)& Shirley \\ 
 hydroxylated& O 1s  &  In$_2$O$_3$ & 530.1 &1.1&SGL(20) &Linear \\
 & O 1s  &  OH$_\mathrm{ads}$ & 531.4 &1.1&SGL(20) &Linear \\
& O 1s  &  O$_\mathrm{s}$H & 532.3&1.1&SGL(20) &Linear \\ 
\hline
 & In 3d$_\mathrm{5/2}$  &In$_2$O$_3$&444.6&1.1&SGL(30) & Shirley\\ 
 reduced&In 3d$_\mathrm{5/2}$  &In ad-atoms  & 445.4  &1.1 &SGL(30) & Shirley\\
 & O 1s  & In$_2$O$_3$ & 530.2  &1.2 &LA(1,1.9,200) & Linear\\ 
 \hline
& In 3d$_\mathrm{5/2}$  & In$_2$O$_3$ & 444.7 & 1.2 & SGL(30) & Shirley \\ 
 stoichiometric & O 1s  & In$_2$O$_3$ & 530.2  &1.2 & SGL(20)& Linear\\ 
 & O 1s  &  OH$_\mathrm{ads}$ & 531.4 &1.1&SGL(20) &Linear \\
& O 1s  &  O$_\mathrm{s}$H & 532.3&1.1&SGL(20) &Linear \\ 
\hline
\end{tabular}
\caption{Fitting parameters for the XPS spectra in Figure 2 of the main text.}
\end{table}

\begin{table}[H]
\centering
\begin{tabular}{c cccccc}
Surface  & Core&  Peak& Binding energy & FWHM & Line & Background \\
 termination&  level & label &  [eV] & [eV] & shape& shape\\ 
 \specialrule{2.5pt}{1pt}{1pt}
& O 1s  & In$_2$O$_3$ & 530.2  &1.2 & SGL(20)& Linear\\ 
 stoichiometric & O 1s  &  OH$_\mathrm{ads}$ & 531.4 &1.1&SGL(20) &Linear \\
& O 1s  &  O$_\mathrm{s}$H & 532.3&1.1&SGL(20) &Linear \\ 
\hline
& O 1s  & In$_2$O$_3$ & 530.2  &1.1 & SGL(20)& Linear\\ 
 methanol & O 1s  &  O--CH$_\mathrm{3}$ & 531.2 & 1.3 &SGL(20) &Linear \\
& O 1s  &  O$_\mathrm{s}$H & 532.3& 1.3 &SGL(20) &Linear \\ 
\hline
formic & O 1s  & In$_2$O$_3$ & 530.2  &1.1 & SGL(20)& Linear\\ 
acid & O 1s  &  HCOOH & 532.6 & 1.5 &SGL(20) &Linear \\
\hline
\end{tabular}
\caption{Fitting parameters for the O 1s spectra in Figure 4(a) of the main text.}
\end{table}

\begin{table}[H]
\centering
\begin{tabular}{c cccccc}
Surface  & Core&  Peak& Binding energy & FWHM & Line & Background \\
 termination&  level & label &  [eV] & [eV] & shape& shape\\ 
 \specialrule{2.5pt}{1pt}{1pt}
stoichiometric& In 3d$_\mathrm{5/2}$  & In$_2$O$_3$ & 444.7 & 1.2 & SGL(30) & Shirley \\ 
methanol& In 3d$_\mathrm{5/2}$  &  In$_2$O$_3$ & 444.8 &1.1&LA(1,10,450) & Shirley \\ 
formic acid& In 3d$_\mathrm{5/2}$  & In$_2$O$_3$ & 444.7 & 1.2 &LA(1,3,450)& Shirley \\ 
\hline
\end{tabular}
\caption{Fitting parameters for the In 3d$_\mathrm{5/2}$ spectra in Figure 4(b) of the main text.}
\end{table}

\begin{table}[H]
\centering
\begin{tabular}{c cccccc}
Surface  & Core&  Peak& Binding energy & FWHM & Line & Background \\
 termination&  level & label &  [eV] & [eV] & shape& shape\\ 
 \specialrule{2.5pt}{1pt}{1pt}
methanol& C 1s  &  O--CH$_\mathrm{3}$ & 286.7 & 1.2 & SGL(30) & Linear \\ 
formic acid& C 1s & HCOOH & 289.3 & 1.0 & SGL(30) & Linear \\ 
\hline
\end{tabular}
\caption{Fitting parameters for the C 1s spectra in Figure 4(c) of the main text.}
\end{table}

\begin{table}[H]
\centering
\begin{tabular}{c cccccc}
Surface  & Core&  Peak& Binding energy & FWHM & Line & Background \\
 termination&  level & label &  [eV] & [eV] & shape& shape\\ 
  \specialrule{2.5pt}{1pt}{1pt}
stoichiometric& C 1s  &  physisorbed CO$_\mathrm{2}$ & 291.5 & 1.2 & SGL(30) & Linear \\ 
 & C 1s & CO$_\mathrm{3}$  & 289.5 & 1.3 & SGL(30) & Linear \\ 
\hline
reduced& C 1s  &  physisorbed CO$_\mathrm{2}$ & 291.7 & 1.3 & SGL(30) & Linear \\ 
 & C 1s & CO$_\mathrm{3}$  & 289.7 & 1.4 & SGL(30) & Linear \\ 
\hline
hydroxylated& C 1s  &  physisorbed CO$_\mathrm{2}$ & 291.5 & 1.2 & SGL(30) & Linear \\ 
 & C 1s & CO$_\mathrm{3}$  & 289.5 & 1.3 & SGL(30) & Linear \\ 
\hline
\end{tabular}
\caption{Fitting parameters for the C 1s spectra in Figure 5(a), 5(b) and 5(c) of the main text.}
\end{table}

\begin{table}[H]
\centering
\begin{tabular}{c cccccc}
Surface  & Core &  Peak& Binding energy & FWHM & Line & Background \\
termination&  level & label &  [eV] & [eV] & shape & shape\\ 
 \specialrule{2.5pt}{1pt}{1pt}
& O 1s  & In$_2$O$_3$ & 530.2  & 1.3 & SGL(20)& Linear\\ 
stoichiometric & O 1s  &  CO$_\mathrm{3}$ & 531.8  &1.1 & SGL(20) & Linear \\
& O 1s  &  physisorbed CO$_\mathrm{2}$ & 535.3 & 1.1 & SGL(20) & Linear \\ 
\hline
& O 1s  & In$_2$O$_3$ & 530.2  & 1.3 & SGL(20)& Linear\\ 
reduced & O 1s  &  CO$_\mathrm{3}$ & 531.7  & 1.6 & SGL(20) & Linear \\
& O 1s  &  physisorbed CO$_\mathrm{2}$ & 535.3 & 1.5 & SGL(20) & Linear \\ 
\hline
& O 1s  & In$_2$O$_3$ & 530.2  & 1.3 & SGL(20)& Linear\\ 
& O 1s  &  OH$_\mathrm{ads}$ & 531.4 & 1.3 & SGL(20) & Linear \\
hydroxylated & O 1s  &  CO$_\mathrm{3}$ & 531.9  & 1.3 & SGL(20) & Linear \\
& O 1s  &  O$_\mathrm{s}$H & 532.4 & 1.3 & SGL(20) & Linear \\ 
& O 1s  &  physisorbed CO$_\mathrm{2}$ & 535.3 & 1.4 & SGL(20) & Linear \\ 
\hline
\end{tabular}
\caption{Fitting parameters for the O 1s spectra in Figure 7 of the main text for the spectra after CO$_\mathrm{2}$ adsorption.}
\end{table}

\begin{table}[H]
\centering
\begin{tabular}{c cccccc}
Surface  & Core&  Peak& Binding energy & FWHM & Line & Background \\
 termination&  level & label  &  [eV] & [eV] & shape& shape\\ 
 \specialrule{2.5pt}{1pt}{1pt}
hydroxylated & In 3d$_\mathrm{5/2}$  & In$_2$O$_3$ & 444.7 & 1.3 & LA(1.5,200)& Shirley \\ 
\hline
reduced & In 3d$_\mathrm{5/2}$  &In$_2$O$_3$&444.6&1.1&SGL(30) & Shirley\\ 
&In 3d$_\mathrm{5/2}$  &In ad-atoms  & 445.3  &1.1 &SGL(30) & Shirley\\
 \hline
stoichiometric & In 3d$_\mathrm{5/2}$  & In$_2$O$_3$ & 444.7 & 1.3 & SGL(30) & Shirley \\ 
\hline
\end{tabular}
\caption{Fitting parameters for the XPS spectra in Figure 8 of the main text for the spectra after CO$_\mathrm{2}$ adsorption.}
\end{table}

\section{LEED images}

\begin{figure}
    \centering
        \includegraphics[width=0.5\textwidth]{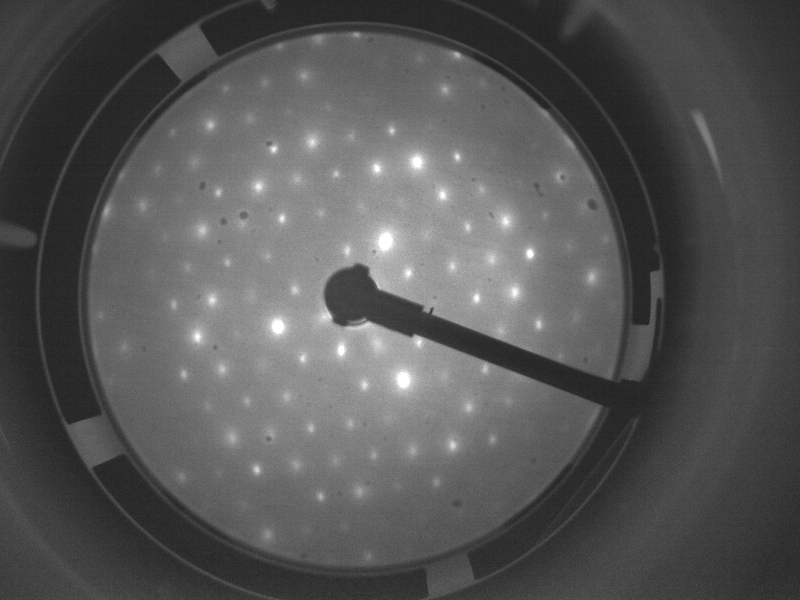}
    \caption{LEED image of the stoichiometric In$_2$O$_3$(111) surface measured with 63 eV.}
    \label{fig:in3dcls}  
\end{figure}

\section{Ad-atoms on reduced In$_2$O$_3$(111)}
The relative stabilities and Bader charges of the In ad-atoms at different sites calculated at the PBE level of theory are given in Table \ref{tab:stabs}. Geometries where the ad-atom binds to the A or C site are considerably higher in energy than at the B site.
 
\begin{table}[htbp]
    \centering
    \begin{tabular}{l c c }
        site & relative stability /eV & Bader charge\\
         \hline
         A     & +0.83  & +0.75 \\
         B$_{\mathrm{b}}$ & +0.00  & +0.75 \\
         B$_{\mathrm{h}}$ &   0    & +0.77 \\
         C     & +1.38  & +0.74 \\
         \hline
    \end{tabular}
    \caption{Stabilities and Bader charges of In ad-atoms on different In$_2$O$_3$(111) high symmetry sites. The stabilities are calculated with respect to B$_{\mathrm{h}}$, the high-symmetry 3-fold hollow site created by O atoms at the B-site. B$_{\mathrm{b}}$ is a bridge site between two of the O atoms. The Bader charges at the different sites are obtained as the difference between the number of Bader electrons and electrons on the neutral atom of the element.}
    \label{tab:stabs}     
\end{table}

\begin{figure}
    \centering
        \includegraphics[width=0.5\textwidth]{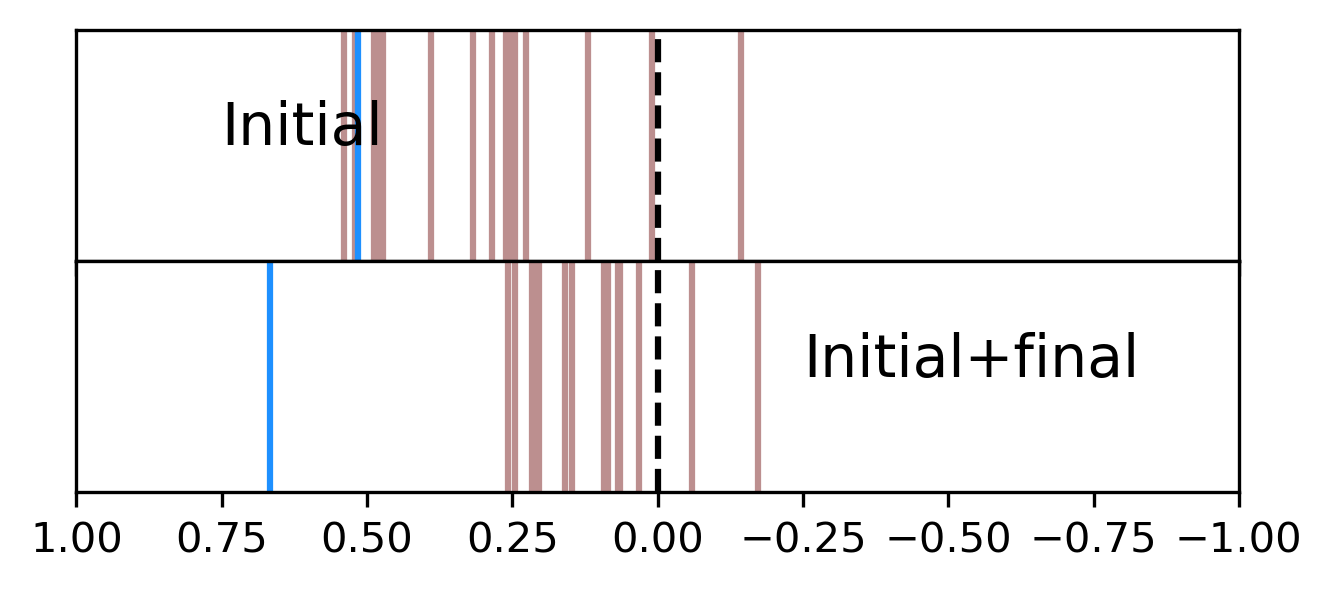}
    \caption{Surface In 3d core-level shifts for an ad-atom at B$_{\mathrm{b}}$ site of the In$_2$O$_3$(111) surface calculated in the electrostatic initial state picture (top) and in the complete screening picture (bottom). The ad-atom CLS are shown in blue, while CLS from other surface In atoms are shown in brown. The dashed line shows the position of the bulk In.}
    \label{fig:invstot}
\end{figure}
The ad-atoms have In 3d core-level shifts (CLS) at higher binding energies with respect to the bulk indium atom (see Figure \ref{fig:in3dcls}). Indeed, the ad-atoms should be shifted to higher binding energies compared to the bulk (and other low-coordinated surface indium atoms), due to the ad-atoms experiencing lower Madelung potentials compared to the bulk.\cite{Nelin2014,Bagus_19} However, final state effects arising from the electronic relaxation upon creating the core hole could still be important. To illustrate this, we obtained the In 3d CLS in the initial state picture by re-calculating the Kohn-Sham eigenvalues of the core states for the ad-atom bound at the B$_{\mathrm{b}}$ site. As can be seen in Fig. \ref{fig:invstot}, the CLS for the ad-atom computed in the initial picture is not distinguishable from the other surface indium atoms. The inclusion of final state effects yields a larger positive shift for the ad-atom and less positive shifts for other surface indium atoms, allowing the shift from the ad-atom to be resolved. 

The In 3d CLS of In atoms in the first O-In-O trilayer were calculated (including final state effects) for all ad-atom containing structures (Fig \ref{fig:in3dcls}). All ad-atoms show a positive shift with respect to an In atom in the centre of the slab, shifted further than any other surface In atom.
\begin{figure}
    \centering
        \includegraphics[width=0.5\textwidth]{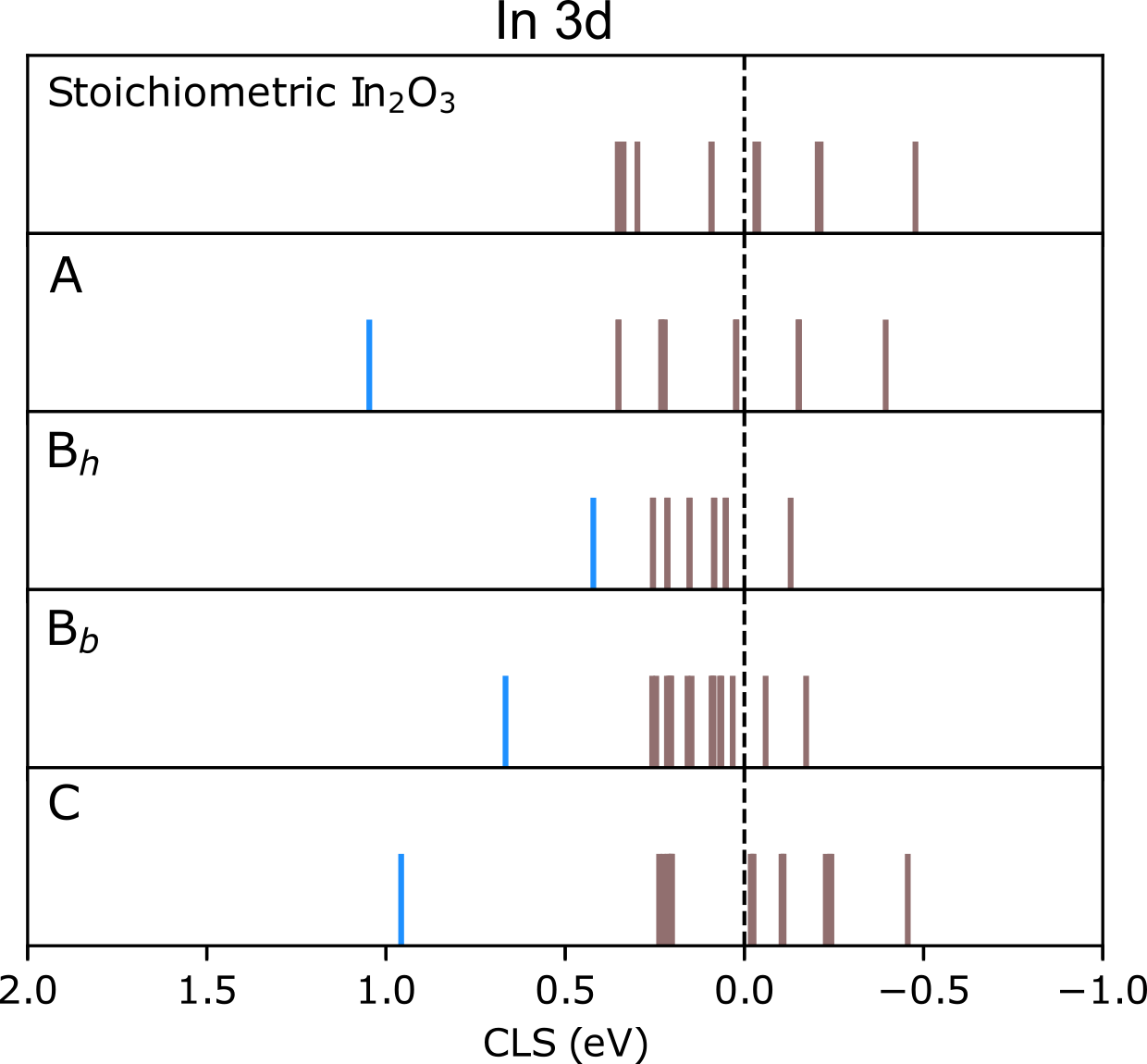}
    \caption{Surface In 3d CLS for stoichiometric and ad-atom containing In$_2$O$_3$(111) surface. The ad-atom CLS are shown in blue, while CLS from other surface In atoms are shown in brown. The dashed line shows the position of the bulk In in each system.}
    \label{fig:in3dcls}  
\end{figure}

\section{Surface coverages of water, formic acid, and methanol}
The effect of coverage on the adsorption energy of water, formic acid, and methanol were investigated on the stoichiometric In$_2$O$_3$(111) in order to determine their saturation coverages. For each adsorbate, the most favourable adsorption sites were determined first for a single molecule in the unit cell. All three molecules prefer to adsorb at the B site dissociatively, formic acid having the strongest binding energy. The next best adsorption site for all molecules is around the C site, with water and methanol adsorbing molecularly to the site, while formic acid preferably donates a proton to a surface oxygen, binding in a tilted configuration. The structures with 1-3 molecules per unit cell were constructed by populating all three equivalent B sites one by one and relaxing the structures. Structures with 4-6 molecules per unit cell were constructed by taking the structure with all B sites occupied, adding molecules to site C one by one, and relaxing the obtained structures. 
\begin{figure}
    \centering
        \includegraphics[width=0.5\textwidth]{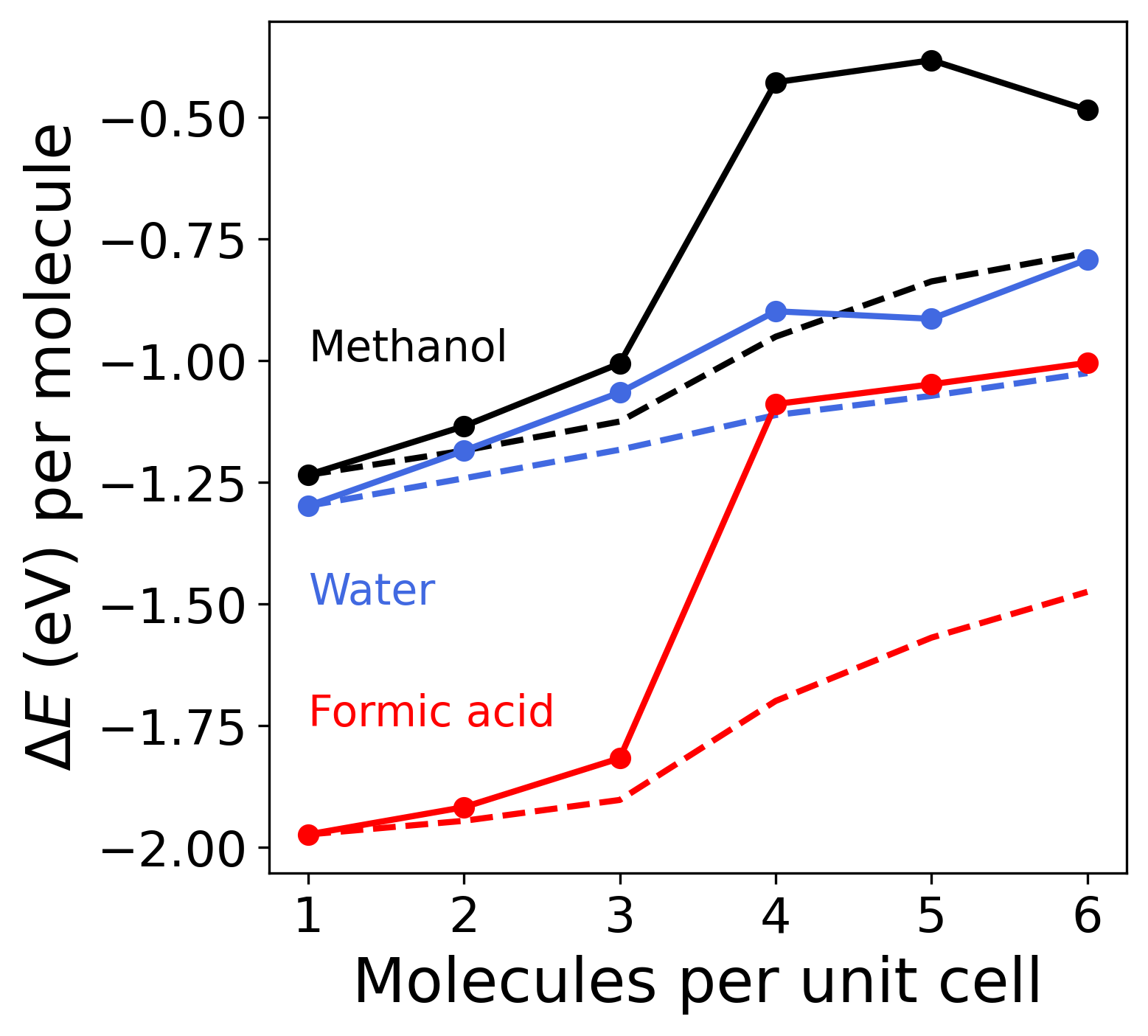}
    \caption{Differential (solid lines) and average (dashed lines) adsorption energies of water (blue), methanol (black), and formic acid (red) as a function of coverage.}
    \label{fig:all_vs_cov}  
\end{figure}
The results (Fig. \ref{fig:all_vs_cov}) show that the adsorption energy to the site B is only slightly lowered when more than one molecule is adsorbed around it. Differential adsorption energy to site C is much lower than site B, especially for formic acid and methanol, with increasing coverage having only a small effect. Based on the adsorption energies, methanol and water should have a saturation coverage of 3 molecules per unit cell at room temperature. Formic acid adsorbs on the surface more strongly, and could achieve a saturation coverage of 6 molecules per unit cell. 

\section{Surface coverage of carbonate on stoichiometric, reduced, and hydroxylated In$_2$O$_3$(111)}
Adsorption of CO$_2$ as a carbonate species was investigated on the stoichiometric, reduced, and hydroxylated terminations of In$_2$O$_3$(111). On the stoichiometric surface, the CO$_2$ molecule binds to the oxygen atom around site B in a bent geometry. Incorporating three CO$_2$ molecules around the site is possible, with the differential adsorption energy decreasing upon addition of the second and third CO$_2$. 

On the reduced surface, the CO$_2$ can adsorb at the B-site oxygen atoms, coordinating also to the ad-atom which is adsorbed at the center of the B site. The adsorption of up to three CO$_2$ molecules around site B on the reduced In$_2$O$_3$(111) is more exothermic than on the stoichimetric In$_2$O$_3$(111), indicating that the presence of the ad-atom does not block adsorption of CO$_2$ as carbonate. 
\begin{figure}
    \centering
        \includegraphics[width=0.5\textwidth]{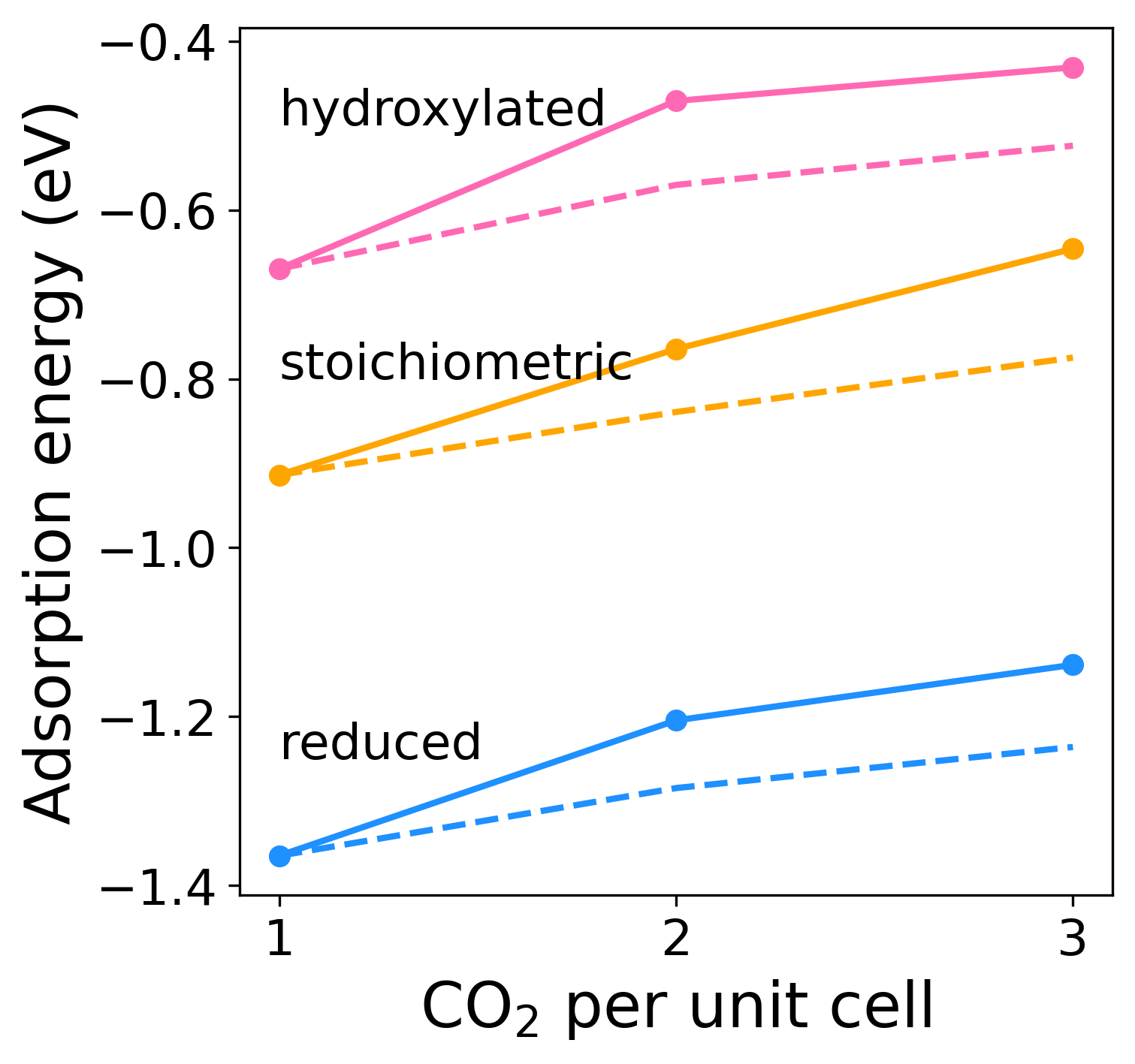}
    \caption{Differential (solid lines) and average (dashed lines) adsorption energies of carbonate as a function of coverage on stoichiometric (yellow), reduced (blue), and hydroxylated (pink) In$_2$O$_3$(111).}
    \label{fig:CO2_vs_cov}  
\end{figure}

\section{Effect of exact exchange and OH coverage on O 1s binding energies}

\begin{figure}
    \centering
        \includegraphics[width=0.8\textwidth]{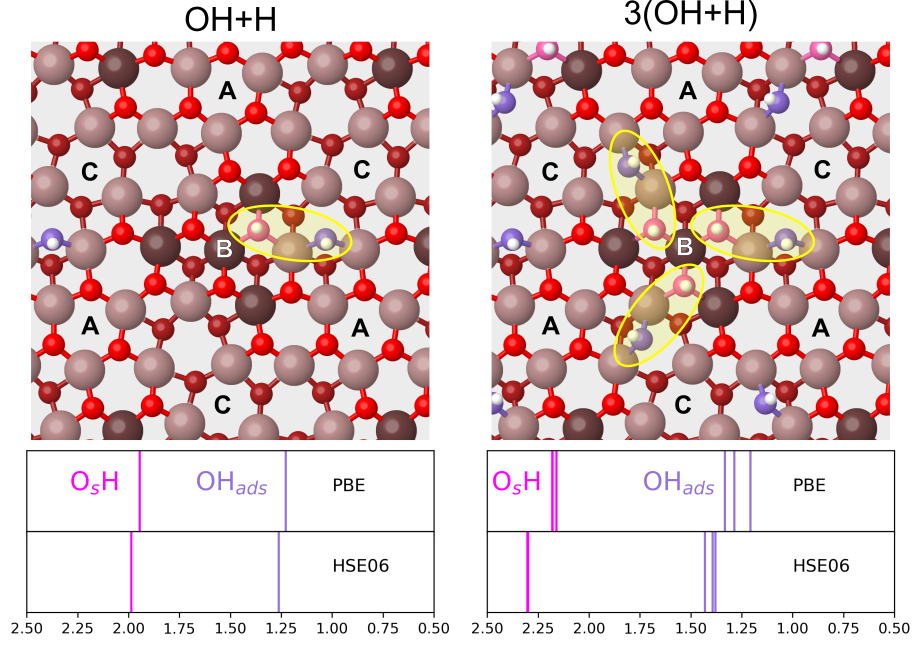}
    \caption{Optimised structures (top) and O 1s CLS (bottom) for hydroxyl group containing In$_2$O$_3$(111) surfaces for two different coverages calculated with PBE and HSE06 functionals.}
    \label{fig:OHcoverage}  
\end{figure}

The O 1s CLS of hydroxyl groups formed on In$_2$O$_3$(111) upon dissociative water adsorption were calculated at the PBE and HSE06 level of theory with one (low coverage) and three (saturation coverage) dissociated water in the unit cell (Fig. \ref{fig:OHcoverage}). At low coverage, the PBE/HSE06 calculated O 1s CLS with respect to a bulk oxygen are 1.23/1.26 and 1.95/1.99 eV for the OH$_{\mathrm{ads}}$ and O$_{\mathrm{s}}$H groups, respectively. At the saturation coverage, the OH$_{\mathrm{ads}}$ and O$_{\mathrm{s}}$H groups give slightly different O 1s CLS, the average CLS computed using PBE/HSE06 being 1.28/1.40 and 2.17/2.30 eV, respectively. The PBE computed values at saturation coverage give the best agreement with experimental binding energy shifts (1.3 and 2.1 eV for OH$_{\mathrm{ads}}$ and O$_{\mathrm{s}}$H, respectively). The PBE and HSE06 computed values at low coverage slightly underestimate the shifts with respect to experiments, especially for the O$_{\mathrm{s}}$H species, while the HSE06 calculated values at saturation coverage slightly overestimate the shifts. 

\section{Carbon containing adsorbates on In$_2$O$_3$(111)}

\begin{figure}
    \centering
        \includegraphics[width=0.8\textwidth]{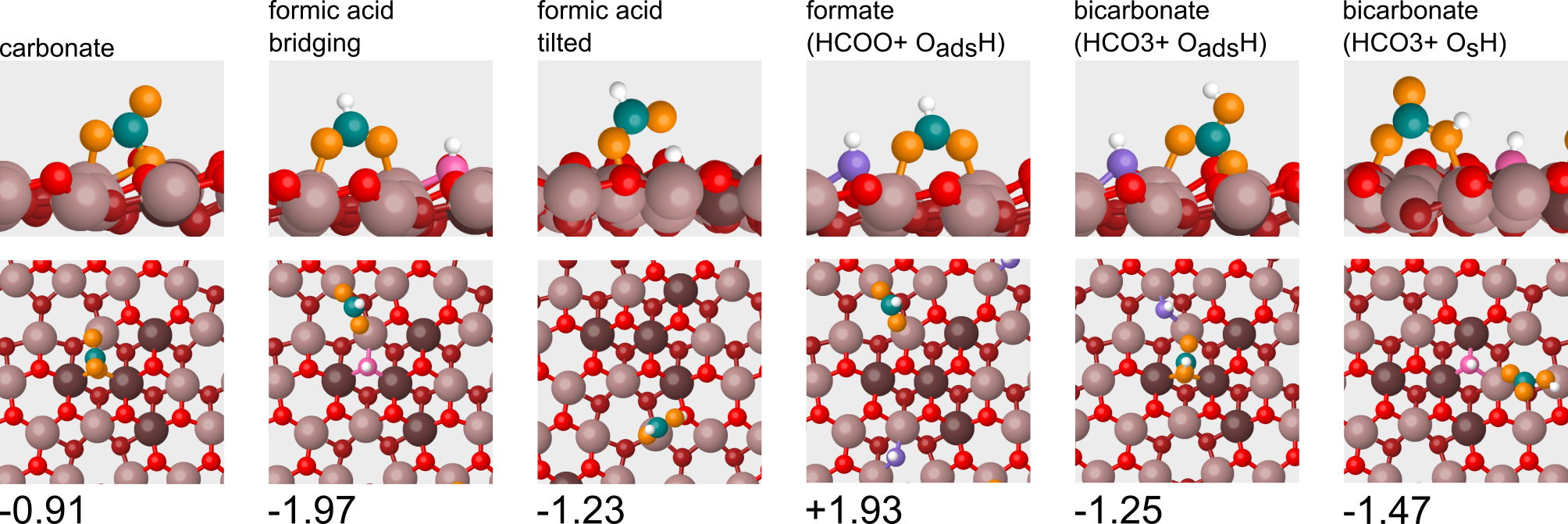}
    \caption{Optimised geometries and adsorption energies (in eV) of various carbon containing adsorbates on In$_2$O$_3$(111). The adsorption energies are given with respect to the clean stoichimetric In$_2$O$_3$(111) slab and CO$_2$ (carbonate), formic acid (formic acid), or CO$_2$ and H$_2$O (formate and bicarbonate) in the gas-phase.}
    \label{fig:carbs}  
\end{figure}
\begin{figure}
    \centering
        \includegraphics[width=0.5\textwidth]{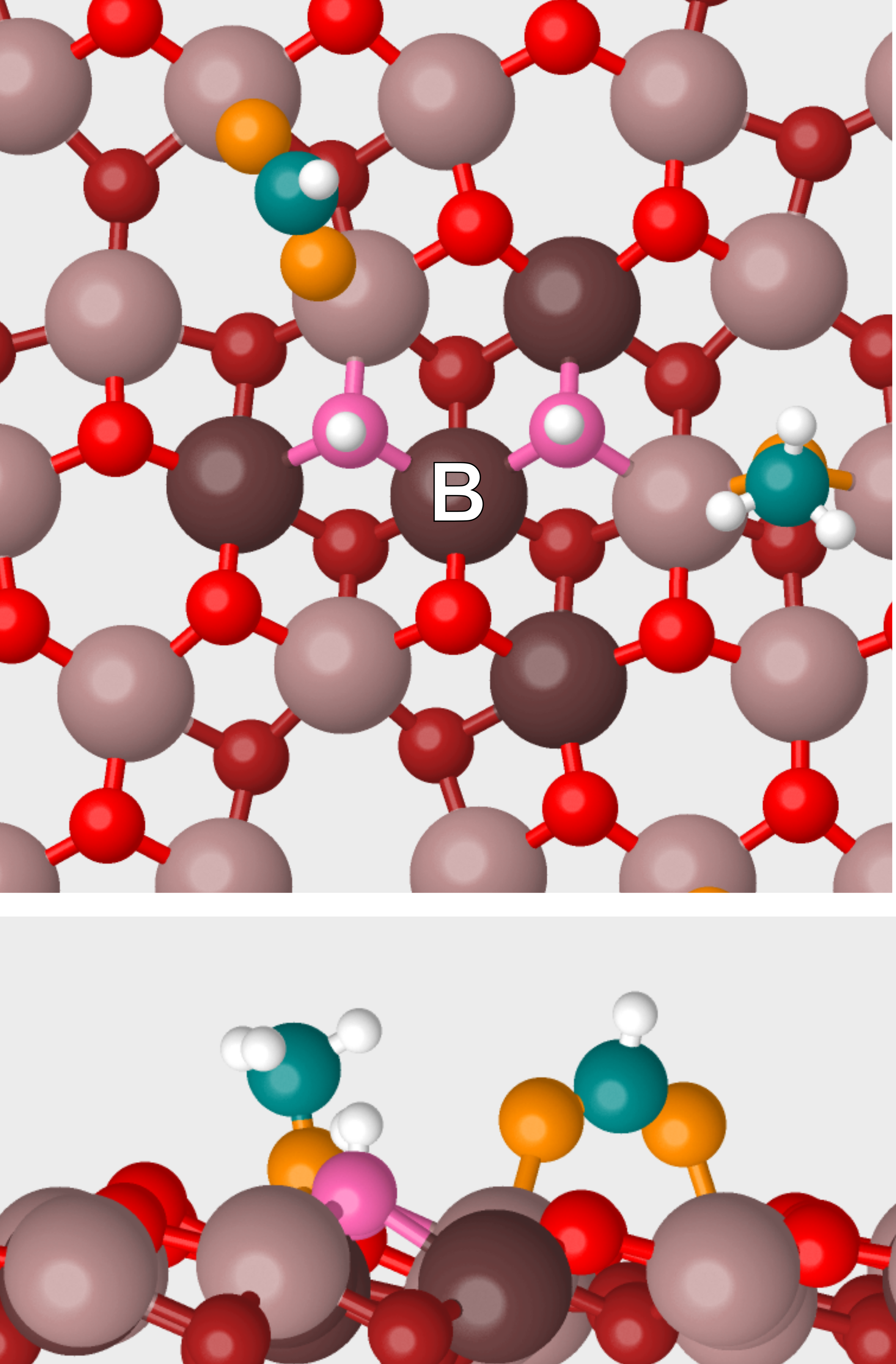}
    \caption{Formic acid and methanol adsorbed in the same In$_2$O$_3$(111) unit cell for C 1s $\Delta$CLS calcualtion.}
    \label{fig:hcooh_meoh}  
\end{figure}

In order to support the peak assignment of the experimental XPS spectra of the CO$_2$ adsorption studies, the C 1s CLS of various possible adsorption structures (Fig. \ref{fig:carbs}) have to be calculated. However, only relative CLS can be obtained within the applied methodology. Since there are no bulk carbon atoms in the stoichiometric In$_2$O$_3$ structure, there is no reference carbon atom in the system, unlike in the case of oxygen and indium. Therefore to provide a reference molecule on the surface that CLS of other adsorbates could be calculated with reference to, the adsorption of formic acid and methanol were studied experimentally on the stoichiometric In$_2$O$_3$(111). Both molecules give clear peaks in the C 1s spectra, separated by 2.6 eV. 
To calculate the relative shift between formic acid and methanol, both molecules were placed on a stoichiometric In$_2$O$_3$(111) slab in the same unit cell (Fig. \ref{fig:hcooh_meoh}). 
The relative C 1s core-level shift ($\Delta$CLS), was then calculated with the gradient-corrected PBE functional as a difference in the total energy of the system with the core hole either on the methanol or on the formic acid molecule. The calculated $\Delta$CLS was $1.42$ eV, clearly not in agreement with the experimental value of $2.6$ eV. 
Previous calculations have shown that non-local exchange effects included in hybrid exchange-correlation functionals can lead to better agreement between the calculated CLS and experimental XPS data.\cite{VdBossche_JCP_14,ESCA} To investigate this effect for CLS of In$_2$O$_3$(111) surface species, the C 1s $\Delta$CLS of various carbon containing adsorbates were calculated using PBE and HSE06 functionals, accompanied by Bader charge analysis. The C 1s $\Delta$CLS and Bader charges are presented in table \ref{tab:cls_comp}.
\begin{table}
\centering
\begin{tabular}{lccccccccc}
 & \multicolumn{3}{c}{$\Delta$CLS (eV)} & \multicolumn{3}{c}{b.c. on C, adsorbate} & \multicolumn{3}{c}{b.c. on C, MeOH} \\
  \hline 
 Structure & PBE & HSE06 & $\Delta$ & PBE & HSE06 & $\Delta$ & PBE & HSE06 & $\Delta$\\ 
  \hline  
 formic acid bridging  & 1.42 & 2.78 & 1.36 & 1.56 & 1.66 & 0.10 & 0.42 & 0.45 & 0.03 \\ 
 formic acid tilted & 2.17 & 2.82 & 0.65 & 1.52 & 1.62 & 0.10 & 0.48 & 0.52 & 0.04 \\  
 carbonate & 2.42 & 2.92 & 0.50 & 2.05 & 2.22 & 0.17 & 0.46 & 0.49 & 0.03 \\
 formate (HCOO+O$_{\mathrm{ads}}$H) & 1.22 & 3.00 & 1.78 & 1.57 & 1.67 & 0.10 & 0.46 & 0.49 & 0.03 \\ 
 bicarbonate (HCO$_3$+O$_{\mathrm{s}}$H) & 3.27 & 3.91 & 0.64 & 2.08 & 2.22 & 0.14 & 0.47 & 0.49 & 0.02 \\
 bicarbonate (HCO$_3$+O$_{\mathrm{ads}}$H) & 3.82 & 4.65 & 0.83 & 2.18 & 2.29 & 0.11 & 0.43 & 0.46 & 0.03 \\
\end{tabular}
\caption{C 1s CLS (relative to methanol) and Bader charges (b.c.) on the carbon atom of carbon-containing adsorbates on stoichiometric In$_2$O$_3$(111) calculated with PBE and HSE06.}
\label{tab:cls_comp}
\end{table}
The HSE06 calculated C 1s $\Delta$CLS for the formic acid is in much better agreement with the experimental value. In general, all carbon containing adsorbates studied here present more positively shifted C 1s $\Delta$CLS when calculated with HSE06 as compared to PBE. This is consistent with the improved charge separation achieved with the HSE06 functional, the effect of which can be seen as more positive Bader charges on the adsorbate carbon atoms. 
\bibliography{refs}